\documentclass[manuscript]{aastex}

\usepackage{graphicx}
\usepackage{natbib}
\usepackage{amsmath}
\usepackage[version=3]{mhchem} 

\shorttitle{Heat transport and atmospheric collapse around red dwarf stars}
\shortauthors{Wordsworth}

\begin{document}


\title{Atmospheric heat redistribution and collapse on tidally locked rocky planets}


\author{Robin Wordsworth\\
\vspace{0.1in}
\normalsize{School of Engineering and Applied Sciences, Harvard University}\\
\normalsize{Cambridge, MA 02138, USA}\\
}

\begin{abstract}
Atmospheric collapse is likely to be of fundamental importance to tidally locked rocky exoplanets but remains understudied.  
Here, general results on the heat transport and stability of tidally locked terrestrial-type atmospheres are reported. First, the problem is modeled with an idealized 3D general circulation model (GCM) with gray gas radiative transfer. It is shown that over a wide range of parameters the atmospheric boundary layer, rather than the large-scale circulation, is the key to understanding the planetary energy balance. Through a scaling analysis of the interhemispheric energy transfer, theoretical expressions for the day-night temperature difference and surface wind speed are created that reproduce the GCM results without tuning. Next, the GCM is used with correlated-$k$ radiative transfer to study heat transport for two real gases (\ce{CO2} and \ce{CO}). For \ce{CO2}, empirical formulae for the collapse pressure as a function of planetary mass and stellar flux are  produced, and critical pressures for atmospheric collapse at Earth's stellar flux are obtained that are around five times higher (0.14~bar) than previous gray gas estimates. These results provide constraints on atmospheric stability that will aid in future interpretation of observations and exoplanet habitability modeling.
\end{abstract}

\maketitle

\section{Introduction}

M-class (red dwarf) stars comprise around 75\% of the total stellar population of the galaxy \citep{Reid2000}, and are attractive targets in searches for nearby terrestrial planets because of the well-known scaling of transit probability and radial velocity signal with stellar mass and planetary orbit \citep{Seager2010}. 
Ongoing radial velocity and transit surveys have revealed a number of low mass exoplanets and planet candidates in the stellar neighborhood over the last decade \citep{Mayor2009,Charbonneau2009,Pepe2011,Tuomi2012}, and future dedicated missions such as TESS and Plato are likely to discover many more. Furthermore, transit spectroscopy studies of low mass `super-Earth' and `mini-Neptune' planets around M-stars have advanced significantly over the past few years \citep{Bean2010,Croll2011,Demory2012,Kreidberg2013,Fraine2014}. 
Hot, close-in planets around M-stars are likely to be the first rocky planets for which transit spectroscopy and phase curve observations are possible, making them particularly important targets for theoretical study  \citep{Selsis2011,Castan2011,Miguel2011,Samuel2014}. 

Because of these favorable observational conditions, M-stars are also prime targets in the search for habitable Earth-like\footnote{We do not attempt to define the term `Earth-like' here, although as a minimum it usually implies a similar planetary mass and received stellar flux.} planets outside the Solar System. However, if habitable planets around M-stars do exist, they are likely to be very different from Earth. First, M-stars have increased extreme ultra-violet (XUV) emission and coronal mass ejection (CME) for a much longer period than G-class (Sun-like) stars \citep{Khodachenko2007,Lammer2007,Linsky2013}. This means that planets around them may suffer greatly enhanced atmospheric loss, particularly if they lack a magnetic field in the early stages of their evolution \citep{Lammer2007,Tian2009,Cohen2014}. Because the most volatile gases (\ce{H2}, \ce{N2}, \ce{Ar} etc.) will generally degas into a planet's atmosphere rapidly during accretion, they are particularly vulnerable to early loss to space. As a result, wide variations in the initial compositions and total masses of the atmospheres of rocky planets around M-stars should be expected. 

The low relative luminosities of M-stars mean that planets orbiting close enough to receive Earth-like fluxes are likely to be in tidally resonant or entirely locked states. This means that in some cases, the atmospheres retained by tidally rocky planets after the initial stage of loss to space will be unstable to collapse on the surface \citep{Kasting1993}. Despite the fundamental importance of this process, it has to date received relatively little theoretical attention. Previous work has addressed the problem using 3D general circulation models (GCMs) in the context of essentially Earth-like planets \citep{Joshi1997,Joshi2003} and higher mass super-Earths \citep{Wordsworth2011,Selsis2011}. Based on these studies, it was found that the stability of a \ce{CO2} atmosphere is a strong function of a) the total atmospheric mass, b) the stellar flux received by the planet and c) the radiative transfer model used. Using a GCM with simple gray gas radiative transfer and an Earth-like stellar flux $F_{E}=1366$~W~m$^{-2}$, \cite{Joshi1997} found that atmospheres above around 0.03~bar $p_{\ce{CO2}}$ were stable. In contrast, \cite{Wordsworth2011} used a GCM with realistic correlated-$k$ radiative transfer, and found that for a 2.3$r_E$ planet receiving around 30\% of Earth's incident stellar flux, atmospheric collapse could occur for $p_{\ce{CO2}}$ values as large as $10$~bar. 

A comprehensive study of parameter space for the general collapse problem has not yet been performed. \cite{Castan2011} modeled the vapor-equilibrium atmospheres of extremely hot super-Earths using an approach first developed to study Io \citep{Ingersoll1985}, but did not investigate the transition to uncondensed atmospheres. \cite{Heng2012} used an analytical approach to construct stability diagrams across a range of parameters. However, they did not investigate the role of real gas radiative transfer or the planetary boundary layer. As will be shown here, both of these processes are critical to understanding atmospheric collapse. 

Numerous studies of atmospheric circulation on tidally locked terrestrial planets have also recently been conducted [e.g., \cite{Joshi2003,Merlis2010,Heng2011,Pierrehumbert2011b,Edson2012,Leconte2013,Yang2013}]. These studies have elucidated several dynamical and climatic processes that are likely to be important in the tidally locked regime. However, most of them focused on atmospheres with 1~bar surface pressure and Earth-like composition\footnote{Recently, \cite{Wang2012} and \cite{Kaspi2014} used a GCM to simulate changes in circulation for a range of parameters, including surface pressure. However, they focused on planets with zonally symmetric average insolation patterns, which are much less vulnerable to atmospheric collapse than the cases we study here.}. Because of the uncertainties in volatile delivery and atmospheric erosion, there is no justification in assuming that Earth-mass planets will possess Earth-like atmospheres (or oceans) in general.  Hence study of a wider range of scenarios is necessary.

Understanding atmospheric heat redistribution and collapse is particularly important in the context of future spectral observations of rocky planets. This is true both for transiting planets, and potentially for non-transiting cases if phase curve information can be derived \citep{Selsis2011}. Given the complexity of modern GCMs, however, improvements in modeling accuracy must be balanced by  advances in basic theory if insight from future observations is to be maximized. 

Here a general study of atmospheric stability for tidally locked planets around M-stars is conducted. First, idealized gray-gas GCM simulations are used to study the problem of heat redistribution from day to night side. It is shown that in the physically important limit of slowly rotating optically thin atmospheres, the nightside surface temperature can be estimated analytically, without recourse to any parameters derived from the GCM simulations. Next, the GCM is run in multiband correlated-$k$ mode and used to study collapse across a range of stellar fluxes, atmospheric pressures and planetary masses. We focus on pure \ce{CO2} compositions but also study one case where \ce{CO} is the condensing gas. These gases were chosen because they are common products of volcanic outgassing or thermochemistry following bolide impacts but are not easily destroyed via photolysis\footnote{\ce{CO2} photolyzes to \ce{CO} and \ce{O}, which can lead to mixed \ce{CO}/\ce{O2} atmospheres, but catalytic processes prevent this happening in the atmospheres of Mars and Venus \citep{Yung1999}. The collapse behavior of pure \ce{O2} or \ce{N2} should be broadly similar to that of \ce{CO}, except that for these gases, collision-induced absorption is the only source of infrared opacity \citep{Frommhold2006}.} or lost to space, unlike e.g. \ce{H2O} or \ce{CH4}. 
It is found that the properties of the planetary boundary layer are more important to the day-night temperature difference than details of the large-scale circulation. In addition, it is shown that the radiative properties of the condensing gas are a key determinant of nightside temperature, and hence the collapse pressure. 

In Section~\ref{sec:method}, the method used for the 3D simulations is described. In Section~\ref{sec:results} the idealized gray gas simulations are first presented. Next, a simple theoretical model is used to calculate the nightside surface temperatures produced by the GCM from first principles. Finally, correlated-$k$ GCM results are  presented and used to construct simple empirical formulae for the onset of atmospheric collapse  for \ce{CO2} and \ce{CO}  as a function of stellar flux, total atmospheric pressure and planetary mass. In Section~\ref{sec:discuss} the broader implications of the results are discussed and suggestions for future work are given.

\section{Method}\label{sec:method}

For the 3D GCM simulations, the LMD Generic Model is used [e.g., \cite{Wordsworth2011,Wordsworth2013,Leconte2013}]. The key model parameters used are shown in Table~\ref{tab:params}. The LMD model solves the primitive equations on the sphere using the finite-difference approach, with a correlated-k approach for the radiative transfer \citep{Goody1989,Wordsworth2010}. High-resolution line absorption data for input to the correlated-$k$ model was produced using the open-source software \emph{kspectrum} and the HITRAN line database \citep{Rothman2009,Rothman2010}. CO$_2$ collision-induced absorption was included using the GBB parameterisation \citep{Wordsworth2010,Gruszka1998,Baranov2004}, with extrapolation of the data used at temperatures above 400~K. In all cases the surface topography is assumed to be flat. All simulations were performed assuming a single-component ideal gas atmosphere, with atmospheric condensation represented as in \cite{Wordsworth2011}. 
For \ce{CO}, the vapor-pressure curve was computed using a Clausius-Clayperon ideal gas relation, with parameters derived from \cite{CRC2000}. Surface horizontal heat transport (due to e.g., the presence of an ocean) is also neglected.
Both simplifications are chosen to make the problem more tractable, but they also allow a conservative upper limit on the  critical collapse pressure\footnote{Neglecting non-condensing background gases is conservative primarily because pressure broadening of absorption lines increases the atmospheric infrared opacity, but around M-stars increases in the amount of Rayleigh scattering have little effect on planetary albedo [e.g., \cite{Wordsworth2010b,vonParis2010}.]}. Unlike in \cite{Wordsworth2011}, cloud radiative forcing is also neglected. The likely effects of cloud and aerosol radiative forcing are discussed in Section~\ref{sec:discuss}. The surface albedo $A$ is taken to be 0.2, a representative value for rocky planets. For the M-star properties, including the stellar spectrum, data for AD~Leo (Gliese~388) is used, as in previous studies \citep{Segura2003,Wordsworth2010b}. AD~Leo is highly active in the XUV and undergoes frequent flaring events \citep{Shkolnik2009}, but these properties do not concern us here as we are focused on radiative processes in the lower atmosphere. Rayleigh scattering is included as in \cite{Wordsworth2010}, although its effects are limited around M-stars because of their red-shifted spectra. The planetary orbit is assumed circular and the obliquity is set to zero. 
In all simulations the model was run until thermal equilibrium was reached. Complete tidal locking is assumed, allowing the rotation rate $\Omega$ to be calculated from orbital distance via Kepler's third law as
\begin{equation}
\Omega = \frac{2\pi}{1 \mbox{ year}} \left( \frac{M}{M_\odot} \right)^{1\slash 2}\left( \frac{L}{L_\odot} \right)^{-3\slash 4}\left( \frac{F}{F_E} \right)^{3\slash 4} \label{eq:Omega}
\end{equation}
Here $M$, $L$ and $F$, are stellar mass, stellar luminosity and stellar flux incident on the planet, with $M_\odot$, $L_\odot$ and $F_E$ solar and terrestrial values, respectively. Given $L=0.024 L_\odot$ and $M = 0.4 M_\odot$ \citep{Pettersen1981,Reiners2009}, (\ref{eq:Omega}) can be written as
\begin{equation}
\Omega = 9.2 \times 10^{-9} F^{3\slash 4}.
\end{equation}
Hence even a planet close enough to receive a stellar flux of $2 F_E$ (2732~W~m$^{-2}$) around a typical M-dwarf rotates 20~times less rapidly than Earth. This is the key reason why the effects of rotation on atmospheric circulation are far less important for Earth-like planets around M-stars than for Earth.

For the planetary boundary layer (PBL), the Mellor-Yamada  parameterization is used \citep{Mellor1982}, with the modifications proposed by  \cite{Galperin1988} included. This scheme is known to perform well under most conditions in the Martian atmosphere \citep{Haberle1993,Forget1999}, which has several similarities to the cases studied here. 
In brief, the Mellor-Yamada / Galperin (MYG) scheme represents turbulent exchange of momentum and heat between the atmosphere and surface using a second-order closure for the Reynolds-averaged equations. Turbulent kinetic energy $q$ is calculated prognostically assuming a balance between production (via either buoyancy-driven convection or shear in the large-scale flow) and dissipation. Vertical  exchange coefficients in the atmosphere and bulk exchange coefficients at the surface are then calculated using $q$, the potential temperature gradient $\partial \Theta \slash \partial z$, the turbulence mixing length $l$ and empirical dimensionless constants derived from experimental data \citep{Mellor1982}.  The turbulence mixing length is calculated from the empirical scaling law of \cite{Blackadar1962} 
\begin{equation}
l = \frac{l_0 \mathcal K z}{\mathcal K z + l_0}
\end{equation}
where $\mathcal K = 0.4$ is the von K\'arm\'an constant and $l_0$ is the maximum attainable mixing length in the boundary layer. 

Turbulent exchange coefficients are calculated for every grid point at every time step. Then, the vertical atmospheric diffusion of horizontal momentum and potential temperature are calculated as 
\begin{equation}
\left. \frac{\partial \mathbf u}{\partial t}\right|_{turb} = \frac{\partial}{\partial z}\left(K_M \frac{\partial \mathbf u}{\partial z}\right)  \label{eq:MY1}
\end{equation}
\begin{equation}
\left. \frac{\partial \Theta}{\partial t}\right|_{turb} = \frac{\partial}{\partial z}\left(K_H \frac{\partial \Theta}{\partial z}\right)  \label{eq:MY2}
\end{equation}
given time $t$, vertical height $z$, horizontal velocity vector $\mathbf u = (u,v)$, potential temperature $\Theta$ and eddy momentum and heat diffusion coefficients $K_M$ and $K_H$. Diffusion coefficients are calculated using Monin-Obukhov scaling \citep{Garratt1994} in terms of $q$, $l$ and the gradient Richardson number, which is defined as the ratio of potential to kinetic energy in the mean flow 
\begin{equation}
Ri = \frac{g}{\Theta}\frac{\partial \Theta}{\partial z}\left[ \left(\frac{\partial u}{\partial z} \right)^2+  \left(\frac{\partial v}{\partial z} \right)^2\right].
\end{equation}
In discrete form, a bulk Richardson number can also be defined as
\begin{equation}
Ri_B = \frac{g\Delta \Theta \slash \Theta}{|\mathbf u|^2 \slash \Delta z},
\end{equation}
where $\Delta z$ is the height from the surface and $\Delta \Theta$ is the potential temperature difference between the surface and the atmospheric region under consideration.

At the surface, bulk exchange is calculated as 
\begin{equation}
\mathcal F(f) = C_D \rho_a |\mathbf u| (f_a-f_s) \label{eq:MY3}
\end{equation}
where $\mathcal F$ is the flux of quantity $f$, $C_D$ is the bulk drag coefficient and  $\rho_a$ and $|\mathbf u|$ are the atmospheric density and wind speed in the first atmospheric layer, respectively. $f_a$ and $f_s$ are the values of $f$ in the first atmospheric layer and at the surface, respectively; for $f=|\mathbf u|$, $f_s=0$.  Finally, the bulk drag coefficient is calculated as 
\begin{equation}
C_D = \left( \frac{\mathcal K}{ln[ z\slash{z_0}]}\right)^2
\end{equation}
where $z_0$ is the roughness height. $z_0$ is essentially a free parameter in an exoplanet context, although it can be  constrained for terrestrial planets based on solar system radar observations [e.g.,  \citep{Downs1975,Head1985,Rosenburg2011}]. Here we adopt a representative value for rocky surfaces (Table~\ref{tab:params}). Note that the dependence of $C_D$ on $z_0$ is weak,  so even order of magnitude variations in roughness length have only a limited effect on the boundary layer behavior \citep{Wordsworth2011}. 

\section{Results}\label{sec:results}

\subsection{Idealised 3D GCM simulations} \label{subsec:idealGCM}
To begin the investigation, we examine idealized 3D simulations where gray gas radiative transfer is assumed and the atmospheric extinction of incoming starlight is neglected. Three surface pressures are used: 0.01, 0.1 and 1~bar. The gas is assumed to have the thermodynamic properties of \ce{CO2}. Earth radius $r_p=r_E$, gravity $g=g_E$ and incoming stellar flux $F = F_E = 1366$~W~m$^{-2}$ are assumed. With a hemispheric mean approximation for the two-stream radiative transfer, total optical depth is 
\begin{equation}
\tau = \frac{ \kappa p_s}{g\overline{\cos \alpha}}, \label{eq:def_tau} 
\end{equation}
with $p_s$ surface pressure,  $g$ gravity and $\overline{\cos\alpha}=0.5$ the mean cosine of infrared emission angle. We set $\kappa=5.0\times10^{-5}$~m$^2$/kg, yielding $\tau = 0.1$ at 0.1~bar, which places us in an optically thin radiative regime for 2 of the 3 simulations. As will be seen, this is the most relevant limit to study for most (but not all) situations where collapse may occur. 

\begin{figure}[h]
	\begin{center}
		{\includegraphics[width=6.5in]{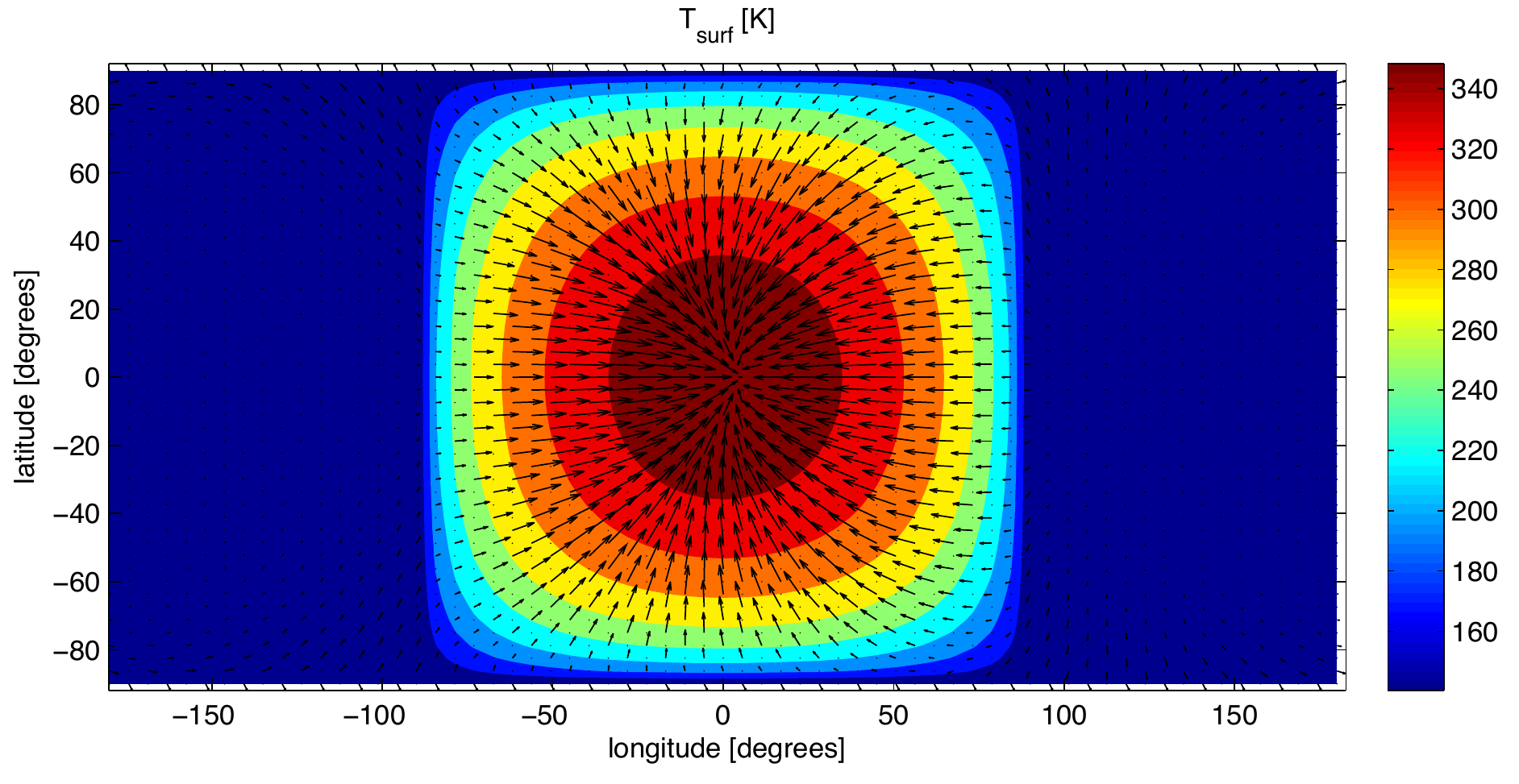}}
	\end{center}
	\caption{Time average surface temperature (color contours) and surface wind (black arrows) in the 0.1~bar gray gas simulation. Some arrows have been removed from the wind field for clarity.}
	\label{fig:Tsurf2D}
\end{figure}

Figure~\ref{fig:Tsurf2D} shows the modeled surface temperature and winds for the intermediate $p_s=0.1$~bar simulation. The picture is a familiar one: a large thermal gradient on the dayside and almost uniform, low temperature on the nightside, with converging surface winds near the substellar point. To a first approximation the flow is longitudinally symmetric, indicating the limited effect of planetary rotation. This is expected based on a dimensional analysis: the ratio of the equatorial Rossby deformation radius\footnote{Note that depending on whether the definition $L_{Ro}=(\sqrt{gH_s}\slash \beta)^{1\slash 2}$ or $L_{Ro}=(\sqrt{N H_s}\slash \beta)^{1\slash 2}$ is used, with $\beta$ the latitudinal gradient in planetary vorticity and $N$ the Brunt-V\"ais\"al\"a frequency, $L_{Ro}$ can be varied by a factor of two. We use the \cite{Leconte2013} definition here to yield a smaller estimate for $L_{Ro}$.} to the planetary radius 
\begin{equation}
\frac{L_{Ro}}{r_p} = \sqrt{\frac{R}{c_p}\frac{\sqrt{c_p T_e }}{2  \Omega r_p} } \label{eq:LecNo}
\end{equation}
is 2.06, indicating a dynamical regime where rotation has little effect on the large-scale circulation \citep{Leconte2013}. In (\ref{eq:LecNo}), $R$ and $c_p$ are the specific gas constant and specific heat capacity at constant pressure (for \ce{CO2} here), $r_p$ is planetary radius and $T_e = [(1-A)F\slash 4 \sigma]^{\frac 14}$ is the global equilibrium temperature. 

\begin{figure}[h]
	\begin{center}
		{\includegraphics[width=6.5in]{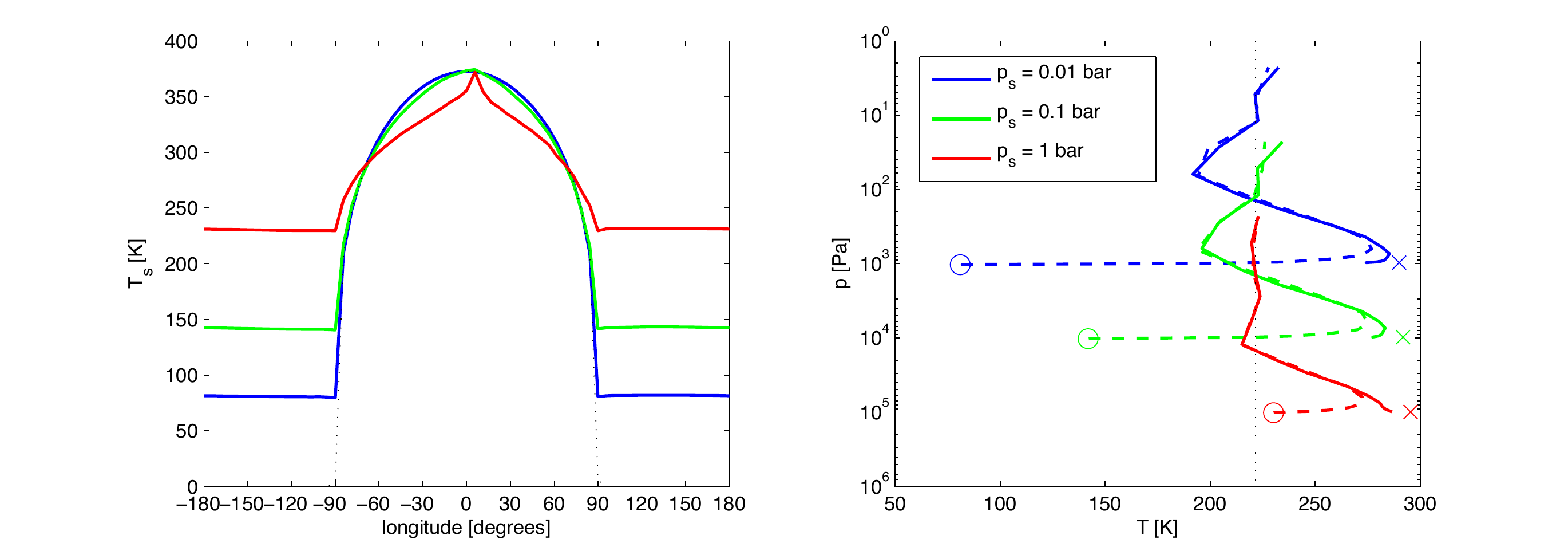}}
	\end{center}
	\caption{(left) Time averaged equatorial surface temperatures at zero latitude in the gray gas simulations. The dotted black line  indicates equilibrium temperature. (right) Hemisphere and time-averaged day and nightside temperatures vs. pressure in the same simulations. Crosses and circles indicate day- and nightside surface temperatures, respectively. The dotted black line indicates  skin temperature.}
	\label{fig:Tprof}
\end{figure}

Figure~\ref{fig:Tprof} (left) shows the surface temperature at the equator for all simulations as a function of longitude. The dayside temperature is extremely close to the local equilibrium temperature (dotted black line) in the 0.01 and 0.1~bar cases, with some deviations occurring at 1~bar. This indicates that the thermal energy transported to the nightside is generally a small fraction of the total --- a fact we will exploit when developing the analytical model in the next section. 

\begin{figure}[h]
	\begin{center}
		{\includegraphics[width=5.0in]{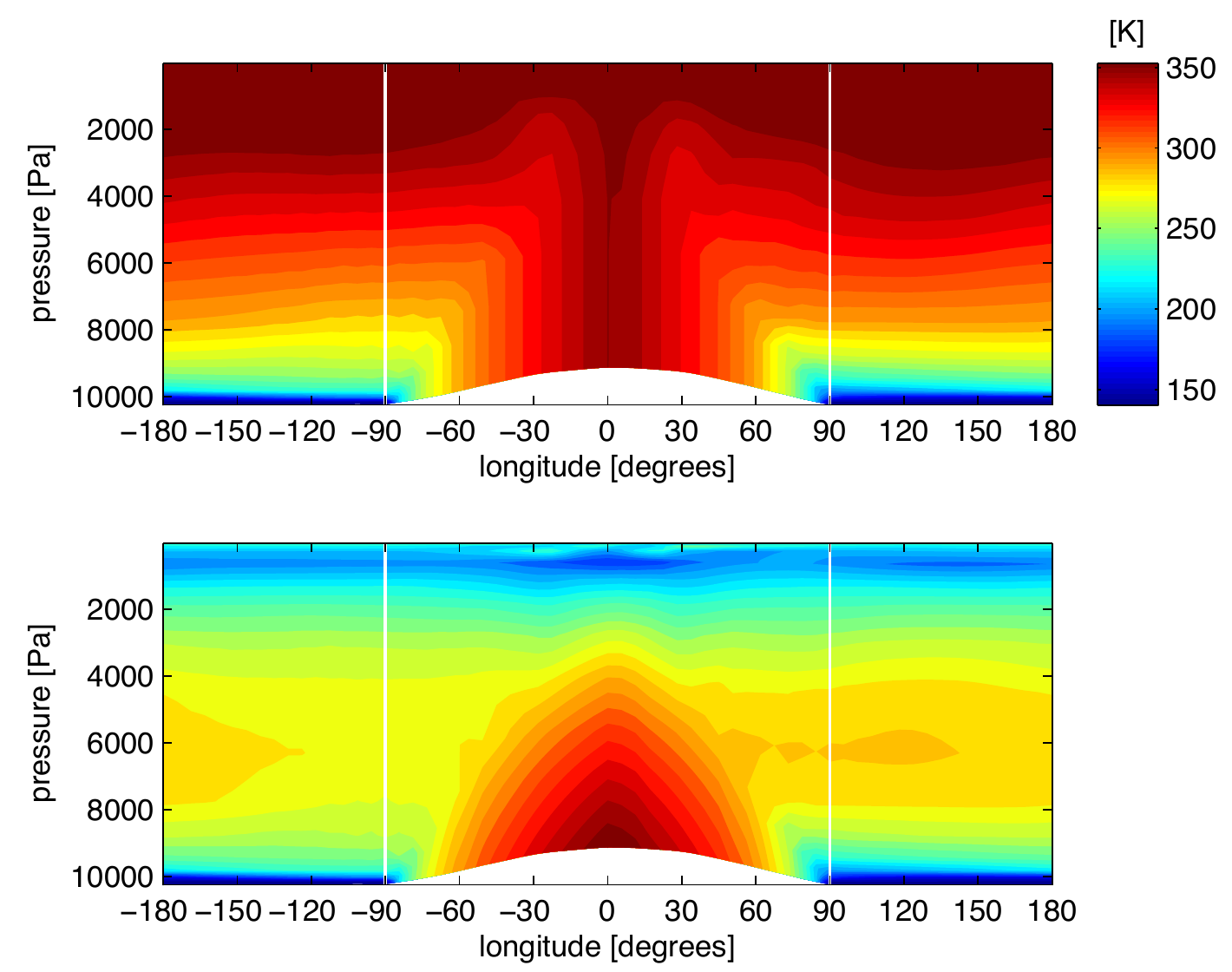}}
	\end{center}
	\caption{Time-averaged longitude-pressure plots of equatorial  potential temperature (top) and temperature (bottom) in the 0.1~bar gray gas simulation. White vertical lines indicate the division between day and night hemispheres.}
	\label{fig:Tslice}
\end{figure}

Figure~\ref{fig:Tprof} (right) shows the day and nightside averaged vertical temperature profiles in the three simulations. All profiles approach the skin temperature $T_{skin}=2^{1/4}T_e$ at low pressures, as is expected given the gray radiative transfer. Remarkably, atmospheric temperatures vary little between day and night hemispheres until extremely close to the surface, despite the large differences in day and night surface temperatures. This general feature of tidally locked terrestrial planet atmospheres, which has been noted previously [e.g., \citep{Merlis2010,Pierrehumbert2011b}], indicates that they are in a weak temperature gradient (WTG) regime \citep{Sobel2001}. This can also be seen from Figure~\ref{fig:Tslice}, which shows longitude-pressure plots of mean atmospheric temperature and potential temperature for the $p_s=0.1$~bar case. Potential temperature $\Theta$ is related to temperature $T$ as $\Theta = T (p_s \slash p)^{R\slash c_p}$, where $p$ and $p_s$ are the atmospheric and surface pressure.

\begin{figure}[h]
	\begin{center}
		{\includegraphics[width=6.5in]{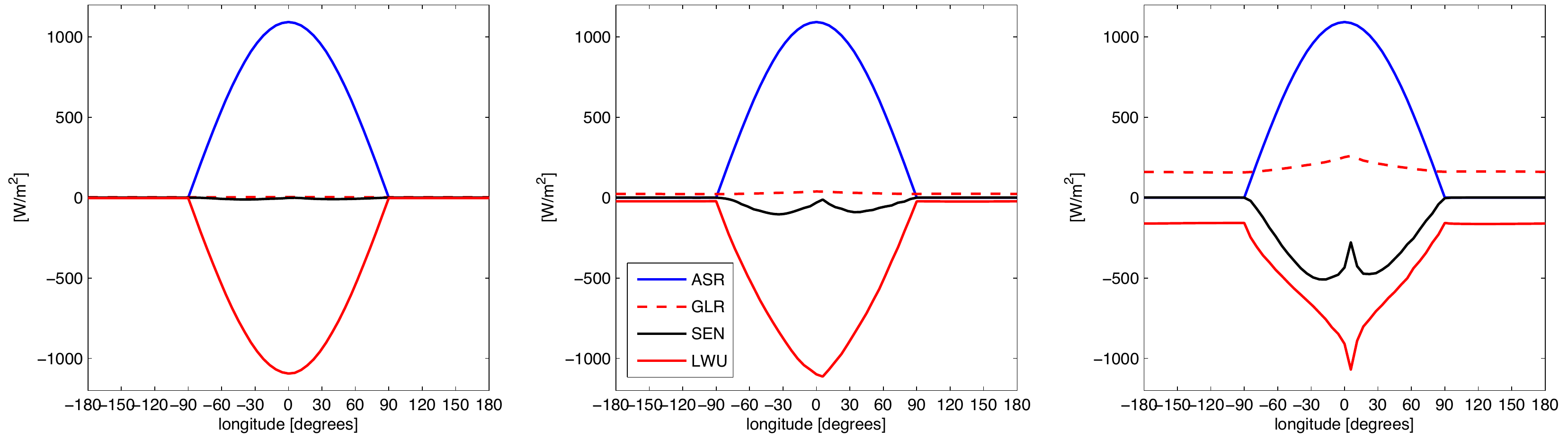}}
	\end{center}
	\caption{Time averaged equatorial surface energy balance at zero latitude in the gray gas simulations. ASR, GLR, SEN and LWU represent absorbed stellar radiation, downwelling infrared radiation to the surface, the net sensible heat flux and the upwelling infrared radiation from the surface, respectively.}
	\label{fig:surf_E_bal}
\end{figure}

To gain further insight into the planetary energy balance, we next study the surface sensible and radiative heat balance. Figure~\ref{fig:surf_E_bal} shows a plot of net sensible heat flux, absorbed stellar radiation, and downwards/upwards infrared radiative fluxes at the surface as a function of longitude for all three simulations. The upwards infrared flux closely matches the absorbed stellar radiation in the 0.01 and 0.1~bar simulations.  The downwards infrared radiation is almost constant in all cases, because the atmospheric temperature varies little with longitude. The sensible heat flux is the most interesting: it exhibits a double-peaked structure on the dayside but drops to close to zero on the nightside.

The behavior of the sensible heat flux is a natural consequence of dry boundary layer physics, as captured by the MYG scheme in the model. On the dayside, fluxes are large because the high stellar energy input to the surface creates an intense, permanent convection layer. In this case, the bulk Richardson number $Ri_B$ is extremely low and hence the drag coefficients in (\ref{eq:MY2}-\ref{eq:MY3}) are large, leading to efficient thermal coupling with the surface. The double-peaked structure occurs because of the surface wind pattern, which we explore in more detail shortly. 

On the nightside, the strong stratification created by the temperature inversion inhibits boundary layer turbulence. The only source of turbulent kinetic energy is the weak large-scale circulation. Mixing is strongly inhibited and the sensible heat flux declines to very low values.  A familiar (although less extreme) analogy to this situation is the polar night on Earth. There, radiative cooling to space dominates the surface heat budget, the planetary boundary layer becomes near-laminar and extremely thin, and the magnitude of the sensible heat flux declines to a low value within days \citep{Cerni1984}. 

\begin{figure}[h]
	\begin{center}
		{\includegraphics[width=5.0in]{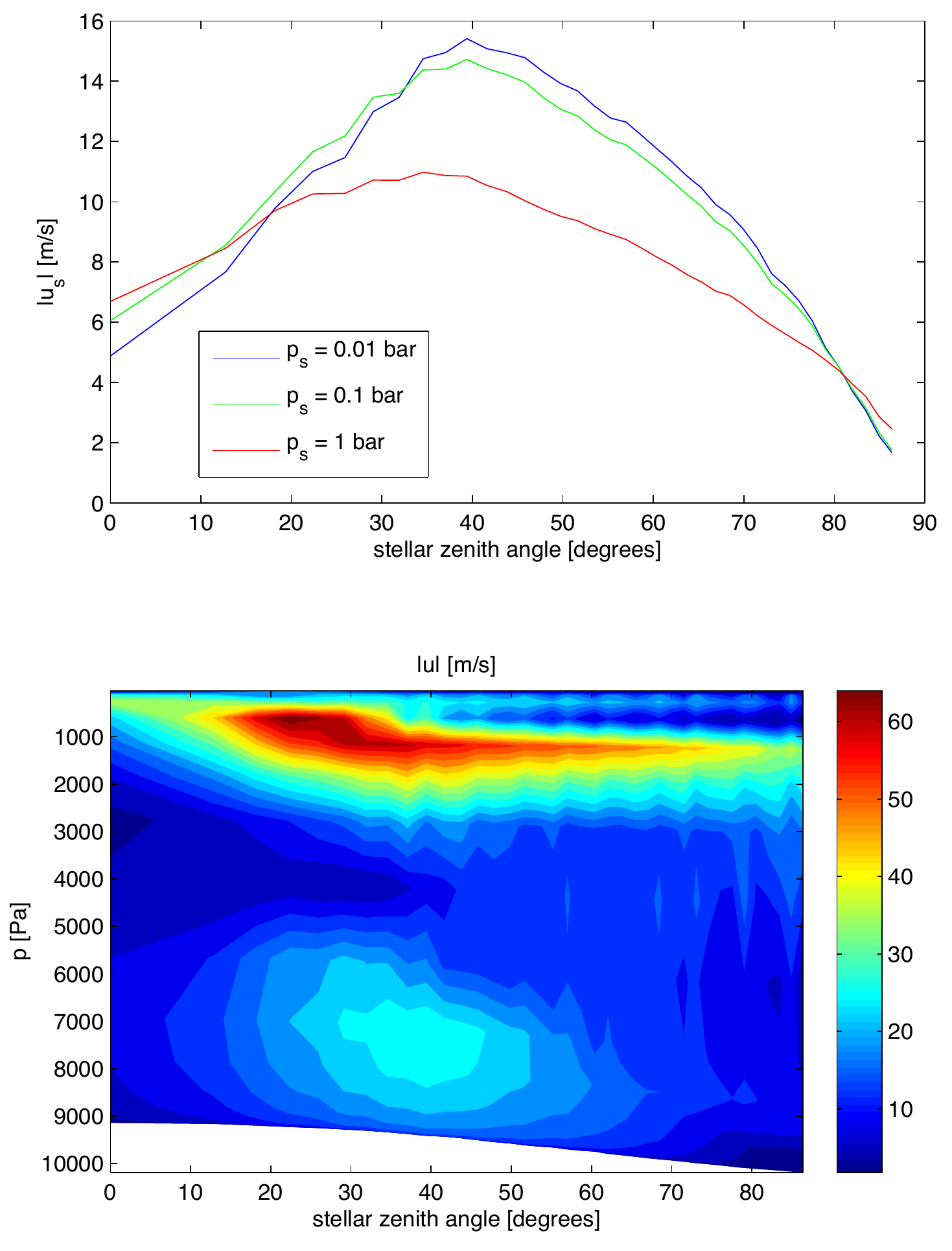}}
	\end{center}
	\caption{(top) Mean surface wind speed vs. stellar zenith angle in the gray gas simulations. (bottom) Mean atmospheric wind speed vs. stellar zenith angle and pressure in the 0.1~bar gray gas simulation.}
	\label{fig:wind_speed}
\end{figure}

Figure~\ref{fig:wind_speed} (top) shows the time-averaged surface wind speed $\theta_z$ in the model as a function of stellar zenith angle. The data in Figure~\ref{fig:wind_speed} was derived by binning the 2D surface velocity magnitude fields $|\mathbf u|=\sqrt{u^2+v^2}$ from the GCM into concentric circles around the substellar point and averaging. All simulations show a peak at around $\theta_z=40^\circ$, with similar profiles at 0.01 and 0.1 bar and a reduced peak wind speed in the 1~bar case. 
The reduction in wind speed near the substellar point explains the dip in the sensible heat flux there (Fig.~\ref{fig:surf_E_bal}) and hence the sharp peak in surface temperature in the 1~bar case. For the 0.01 and 0.1~bar cases, the sensible heat flux is too low to significantly reduce the dayside surface temperatures. 

\begin{figure}[h]
	\begin{center}
		{\includegraphics[width=5.0in]{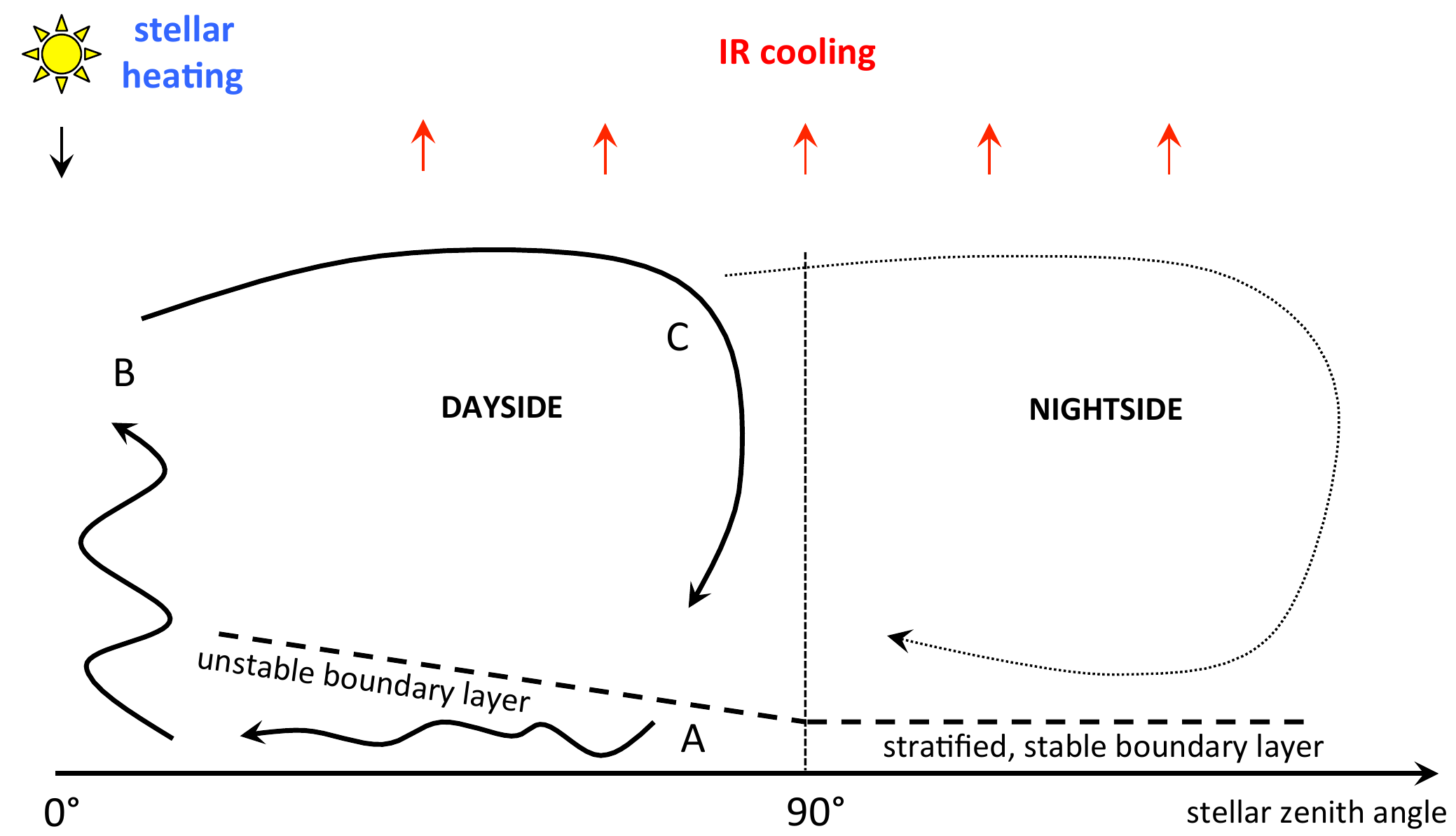}}
	\end{center}
	\caption{Schematic of  key aspects of the circulation and heat transport in a tidally locked atmosphere in contact with a solid surface.}
	\label{fig:schematic}
\end{figure}

The 0.1~bar plot of $|\mathbf u|$ vs. $\theta_z$ and $p$ (bottom) shows an intense jet (up to 60~m~s$^{-1}$) high in the atmosphere. This is the outward branch of the flow seen near the surface in Fig.~\ref{fig:Tsurf2D}; the wind speed minimum around the $p_s/2$ level indicates the transition between flow to/from the substellar point. Essentially, when the planetary rotation rate is low, the large-scale circulation has the form of a single planetary-sized convection cell (see Fig.~\ref{fig:schematic} for a schematic). In the next section, it is shown that the surface wind speed can be explained by a scaling analysis.

It is interesting to compare the 1~bar, $\tau=1$ results in Fig.~\ref{fig:Tprof} (left) with those from the \cite{Joshi1997} study of atmospheric collapse. There, lower day/night side differences were found, with nightside temperatures of around 260-270~K, compared to around 240~K in this study. This suggests that heat redistribution by the atmosphere was much more efficient in their model. The \cite{Joshi1997} model used a simple boundary layer representation based on a quadratic formulation \citep{Joshi1995}, which may have over-represented the efficiency of coupling between the atmosphere and the surface. The subtle issue of accurate sensible heat flux parameterization in the strongly stably stratified regime is discussed further in Section~\ref{sec:discuss}.

\subsection{Analytical model}

Despite the complexity of the GCM used, the results described in the last section appear essentially simple. To gain a deeper understanding, we now reproduce them using a purely analytical approach. Our goal in this section is to calculate the nightside surface temperature $T_{n}$ from first principles, without using any GCM-derived coefficients.

To aid development of the model, a schematic of the key features of the circulation is given in Figure~\ref{fig:schematic}. Inspired by Figure~\ref{fig:Tslice}, we will assume that the atmosphere can be treated as horizontally isothermal outside of the dayside convective zone. Given this, a three-box model of energy exchange between the dayside surface, atmosphere and nightside surface can then be constructed. Our approach is somewhat similar to that of \cite{Pierrehumbert2011b} and \cite{Yang2014}, except here we are after analytical results and hence avoid ad-hoc tuning to the GCM simulations.

Given a dry, single-component atmosphere that is transparent in the visible and gray in the infrared and surface of emissivity equal to 1, the local surface energy balance may be written
\begin{equation}
\sigma T_s^4 =  (1-A)S(\psi,\lambda) + GLR + C_D c_p \rho_a |\mathbf u| (T_a - T_s).\label{eq:surf_ebal}
\end{equation}
Here $S(\psi,\lambda)=F\cos\theta_z$ is the local stellar flux, $\psi$, $\lambda$ and $\theta_z$ are longitude, latitude and stellar zenith angle, respectively, $GLR$ is the downwards infrared radiative flux and $T_a$ is the near-surface atmospheric temperature. Similarly, in the absence of visible absorption the vertically integrated local atmospheric energy balance may be written
\begin{equation}
OAR + GLR = \mathcal A \sigma T_s^4 +  \mathcal D + C_D c_p \rho_a |\mathbf u| (T_s - T_a) \label{eq:atm_ebal}
\end{equation}
where $OAR$ is the outgoing infrared radiation emitted to space by the atmosphere and $\mathcal A$ is the frequency-averaged atmospheric absorptance in the infrared. $\mathcal D$ represents the effects of dynamical transport and will disappear once we assume the atmosphere to be isothermal and perform horizontal averaging. The third term on the right hand side of (\ref{eq:atm_ebal}) is the sensible heat flux: it is simply (\ref{eq:MY3}) with $f=c_pT$ the quantity undergoing exchange.

Next, we define
\begin{equation}
B_d = \frac{\int_{d} \sigma T_s^4 dA }{2\pi r_p^2}
\end{equation}
\begin{equation}
B_n = \frac{\int_{n} \sigma T_s^4 dA }{2\pi r_p^2}
\end{equation}
 with $\int_d dA$ and $\int_n dA$ surface integrals over the planet's day and nightsides, respectively. Applying these integrals separately to (\ref{eq:surf_ebal})  in turn and averaging (\ref{eq:atm_ebal}) over the entire planet, we obtain
\begin{eqnarray}
B_d &=& \frac 12 (1-A)F + GLR + C_D c_p \rho_a \overline{ |\mathbf u| (T_a - T_d) }\label{eq:day_gen} \\
B_n &=& GLR + C_D c_p \rho_a \overline{ |\mathbf u| (T_a - T_n) }\label{eq:night_gen}
\end{eqnarray}
and
\begin{equation}
OAR + GLR = \frac 12 \mathcal A B_d + \frac 12 \mathcal A B_n  + \frac 12 C_D c_p \rho_a \overline{ |\mathbf u| (T_d - T_a)} + \frac 12 C_D c_p \rho_a \overline{|\mathbf u| (T_n - T_a)  }\label{eq:atm_gen}
\end{equation}
where $T_d$ and $T_n$ are the mean dayside and nightside surface temperatures, respectively. 
Note that because of the isothermal assumption, most global average and local atmospheric values are the same [e.g., $\overline{OAR} = (4\pi r_p^2)^{-1}\int_\Sigma OAR dA=OAR$]. We allow for the possibility of horizontal variations in $|\mathbf u|$ and $T_d$, however, by retaining the overbar for the sensible heat terms.

If the atmosphere can be assumed to be optically thin, $\mathcal A \approx \tau$ and $OAR \approx GLR \approx \tau B_a$ \citep{Pierrehumbert2011BOOK}. Then (\ref{eq:day_gen}-\ref{eq:atm_gen}) can be simplified to 
\begin{eqnarray}
B_d &=& \frac 12 (1-A)F + \tau B_a + C_D c_p \rho_a \overline{ |\mathbf u| (T_a - T_d)} \label{eq:day_thin} \\
B_n &=& \tau B_a + C_D c_p \rho_a \overline{|\mathbf u| (T_a - T_n)} \label{eq:night_thin} \\
4  B_a  &=&  B_d  + B_n + C_D c_p \rho_a \overline{ |\mathbf u|\left[ (T_d - T_a) + (T_n - T_a) \right]}\slash \tau
\label{eq:atm_thin}
\end{eqnarray}

\subsection{Purely radiative case}

First we examine the artificial but instructive limit where $C_D\to 0$ and the only permitted surface-atmosphere energy exchange is radiative. Remaining in the optically thin regime, (\ref{eq:day_thin}-\ref{eq:atm_thin}) reduce to 
\begin{eqnarray}
B_d &=& \frac 12 (1-A)F + \tau B_a \label{WTG_1b} \\
B_n &=&  \tau B_a \label{WTG_1a} \\
4  B_a  &=& B_d  + B_n \label{WTG_1c}
\end{eqnarray}
Hence
\begin{equation}
B_n = \frac{\tau}{2-\tau}\frac{(1-A)F}{4} .
\end{equation}
Because we have already assumed $\tau<1$, this may be further approximated as  
\begin{equation}
B_n  \approx \frac{\tau (1-A) F}{8}.
\end{equation}
Hence using the definition of optical depth (\ref{eq:def_tau}) with $\overline{\cos \alpha}=0.5$, 
\begin{equation}
T_n \approx \left(\frac{(1-A)F\kappa p_s}{4 \sigma g}\right)^{1\slash 4}.\label{eq:Tn_simplest}
\end{equation}
In the thin, isothermal limit with zero sensible heat fluxes, this solves the problem of the surface nightside temperature for a given atmospheric opacity and stellar flux. Note that (\ref{eq:Tn_simplest}) can also be derived by neglecting a) the radiative back-reaction of the atmosphere on the dayside surface and b) the back-reaction of the nightside surface on the atmosphere. In other words, it relies on the assumption that the dayside temperature is unaffected by the presence of an atmosphere, and the atmospheric temperature is unaffected by the heat received from the nightside.

Sticking in the numbers to (\ref{eq:Tn_simplest}) for the three ideal GCM cases presented in Section~\ref{subsec:idealGCM}, we get $T_n=$ 70.4, 125.2 and 222.7~K for $p=$ 0.01, 0.1 and 1~bar, respectively. These values are reasonably close to those calculated by the GCM [see Fig.~\ref{fig:Tprof} (left)], which is impressive given the simplicity of the derivation. However, there is a systematic bias towards lower temperatures, which is expected given that we are neglecting all turbulent surface-atmosphere heat exchange. We henceforth refer to (\ref{eq:Tn_simplest}), which provides a lower limit to $T_n$ for optically thin atmospheres, as the \emph{thin radiator temperature}.

\subsection{Inclusion of the dayside sensible heat flux}

Now we relax the  assumption of zero sensible heat fluxes. It might be tempting to include all terms and attempt to solve (\ref{eq:day_thin}-\ref{eq:atm_thin}) numerically. However, a little more physical insight will allow these equations to be simplified further. Specifically, we invoke the fact that radiative fluxes dominate sensible fluxes in the strongly stratified nightside boundary layer (see Fig.~\ref{fig:surf_E_bal}) and drop terms involving $(T_a-T_n)$ in (\ref{eq:night_thin}) and (\ref{eq:atm_thin}). We also assume $B_n<<B_a$ and $\tau B_a<<\frac 12 (1-A)F$. Hence (\ref{eq:day_thin}-\ref{eq:atm_thin}) become 
\begin{eqnarray}
B_d &\approx& \frac 12 (1-A)F  \label{eq:day_thin2} \\
B_n &\approx&  \tau B_a  \label{eq:night_thin2}\\
4  B_a  &\approx&  B_d +  C_D g c_p \overline{ |\mathbf u| (T_d\slash T_a - 1)} \slash (2\kappa R). \label{eq:atm_thin2}
\end{eqnarray}
where the ideal gas law and (\ref{eq:def_tau}) have been used. We rewrite $\overline{ |\mathbf u| (T_d\slash T_a - 1)}$ as $\chi { |\mathbf u| (T_d\slash T_a - 1)}$, where $|\mathbf u|$, $T_d$ and $T_a$ are now taken to be hemispheric mean values and $\chi$ is a factor that accounts for the fact that temperature peaks at $\theta_z=0$, whereas $|\mathbf u|$ peaks around $40^\circ$. To this level of approximation we may write $T_d \slash T_a \propto \cos\theta_z$ and $|\mathbf u|\propto \sin \theta_z$, yielding $\chi = \int_0^{\pi\slash 2} \cos^2\theta_z \sin\theta_z d\theta_z = 1\slash 3$. 
By defining a dimensionless atmospheric temperature $\tilde T = T_a \slash T_d$ and velocity $\tilde U = |\mathbf u|\slash U_0$, with
\begin{equation}
U_0 = (1-A)F \frac{\kappa }{\chi C_Dg }\frac{R}{c_p}
\end{equation}
we can rewrite (\ref{eq:atm_thin2}) as
\begin{equation}
4 \tilde T^4 - 1 = \tilde U(\tilde T^{-1} -1) \label{eq:atm_dimless}.
\end{equation}
(\ref{eq:atm_dimless}) defines $T_a$ (and hence $T_n$) as a function of $T_d$ if $|\mathbf u|$ is known.

\subsection{An equation for $|\mathbf u|$}

To close equation (\ref{eq:atm_dimless}), we next utilize the WTG approximation. WTG scaling for velocity has been analyzed in a shallow water context for Earth \citep{Sobel2001} and was recently applied to hot Jupiters \citep{Perez2013}, suggesting that it should also be useful here. Nonetheless, the central role of the planetary boundary layer in interhemispheric heat transport on rocky planets means that the approach we take here  is rather different from what has been done previously.

The thermodynamic equation \citep{Vallis2006}
\begin{equation}
\frac{DI}{Dt} + \frac{p}{\rho}\nabla \cdot \mathbf u = \mathcal H
\end{equation}
with $I$ internal energy, $p$ pressure, $\rho$ density, $\mathbf u$ velocity, $D\slash Dt$ the advective derivative and $\mathcal H$ the diabatic heating rate in W/kg may be simplified in the WTG, ideal gas limit to
\begin{equation}
\nabla \cdot \mathbf u = \frac{1}{RT_a} \mathcal H \label{eq:WTGdiv}.
\end{equation} 
The essence of (\ref{eq:WTGdiv}) is that in the absence of time-varying or advective effects, expansion of a fluid column in the bulk atmosphere due to heating must be compensated by outward flow of material to eliminate the horizontal thermal gradient. $ \mathcal H $ represents the net effect of all heating and cooling processes in a given region. We first focus on the local radiative heating of the atmosphere by the ground $\mathcal H_{rad,s}$ and determine the probable scaling. Momentarily replacing the unknown $T_a$ by $T_d$, we can write 
\begin{equation}
\frac {U_1}L \sim \frac{1}{RT_d} \mathcal H_{rad,s}
\end{equation} 
and hence
\begin{equation}
 U_1          \sim \frac{L \kappa (1-A)F}{2RT_d} .
\end{equation}
$L$ is a characteristic horizontal length scale that we take here to be the planetary radius $r_p$. The new velocity scale $U_1$ may be related to $U_0$ by defining the dayside scale height $H_d = RT_d \slash g$, resulting in 
\begin{eqnarray}
 U_1 &\sim&\frac{r_p}{H_d}\frac{ (1-A)F\kappa}{2g} \\
 U_1 &\sim&\tilde L U_0
 \end{eqnarray}
with the dimensionless length 
\begin{equation}
\tilde L = \frac{\chi C_D r_p}{2H_d}\frac{c_p}{R}.
 \end{equation}
 Given a roughness height of $1\times10^{-2}$~m and height of the first model layer $z\sim 10$~m, $C_D = 0.0034$.
For an Earth-mass planet with \ce{CO2} atmosphere, this yields $\tilde L = 2.4$. We shall see that $\tilde L$ gives a measure of the strength of $|\mathbf u|$ and hence of the sensible heat flux from the dayside to the atmosphere.
 
\begin{figure}[h]
	\begin{center}
		{\includegraphics[width=4.0in]{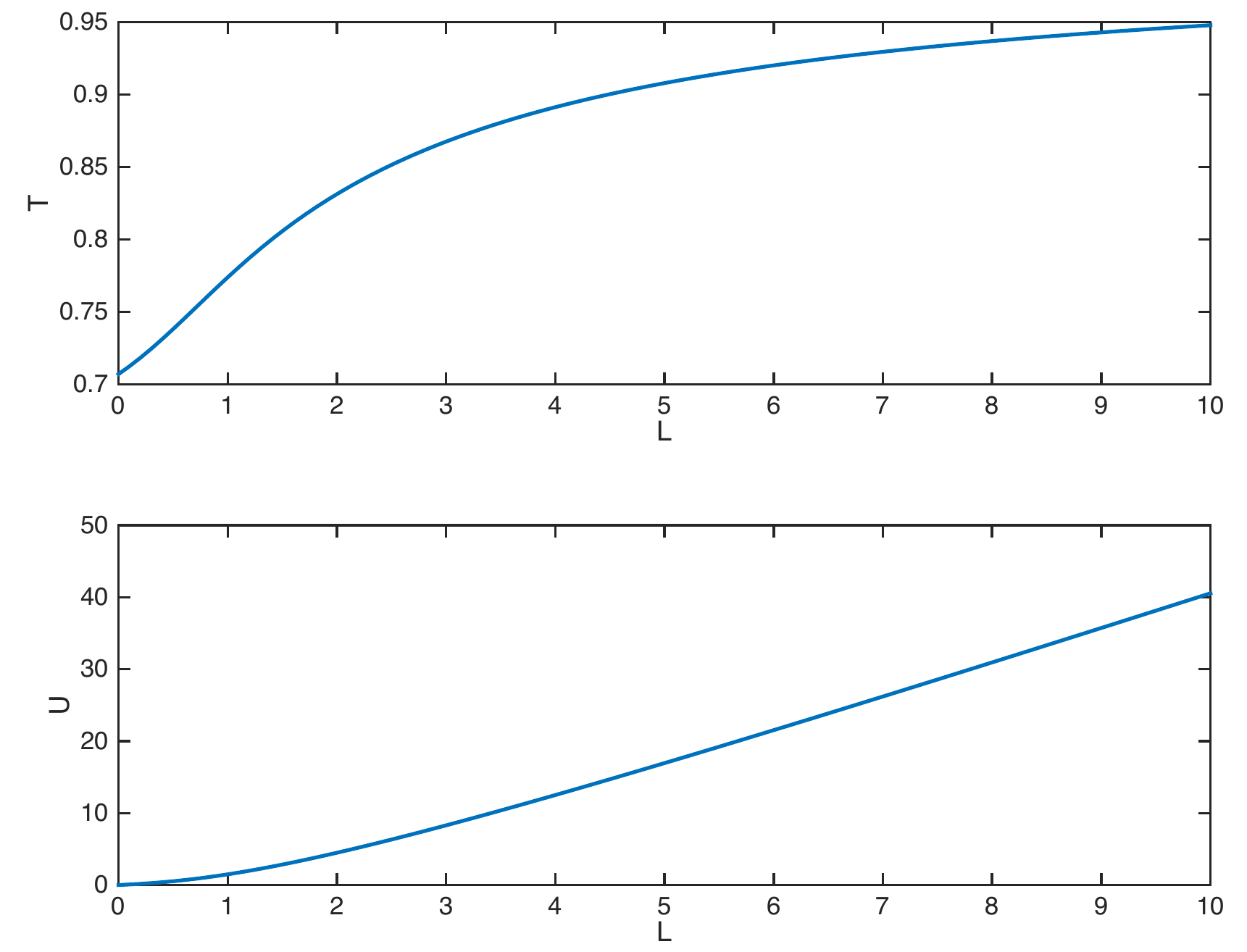}}
	\end{center}
	\caption{Plot of the dimensionless temperature $\tilde T$ (top) and velocity $\tilde U$ (bottom) as a function of the dimensionless length $\tilde L$.}
	\label{fig:scalings_UTL}
\end{figure}

Next we return to (\ref{eq:WTGdiv}), write $\nabla \cdot \mathbf u \sim |\mathbf u|\slash r_p$, and include the dayside radiative cooling and sensible heating terms. This yields 
\begin{equation}
\frac {|\mathbf u|}{r_p} = \frac{1}{RT_a} \left(\kappa \sigma T_d^4 -  2 \kappa \sigma T_a^4 + \frac{g \chi C_D |\mathbf u| c_p(T_d-T_a)}{R T_a}\right) 
\label{eq:WTGdiv2}
\end{equation} 
with the final term on the right hand side the sensible heat flux to the atmosphere in W~m$^{-2}$ divided by a factor $p_s \slash g$ to get the heating rate per unit mass. Using the previously derived scaling relations, (\ref{eq:WTGdiv2}) can be used to create an expression for the dimensionless velocity  
\begin{equation}
\tilde U = \frac{\tilde L(1 -  2 \tilde T^4)}{\tilde T + 2\tilde L  (1 - \tilde T^{-1})}.
\label{eq:WTGdiv_dimless}
\end{equation} 
Substituting (\ref{eq:WTGdiv_dimless}) into (\ref{eq:atm_dimless}) we get
\begin{equation}
(4 \tilde T^4 - 1)\left[\tilde T + 2\tilde L (1 - \tilde T^{-1}) \right] - \tilde L (1 -  2 \tilde T^4)(\tilde T^{-1} -1) = 0.
\end{equation}
This polynomial in $\tilde T$ and $\tilde L$ can be solved by Newton's method, resulting in expressions for $\tilde T$ and $\tilde U$ as functions of $\tilde L$, which is a fixed parameter for a given planet. The result is shown in Fig.~\ref{fig:scalings_UTL}. Given $\tilde L=2.4$, we find $\tilde U=6.2$ and $\tilde T=0.85$. For $U_0=1.2$~m~s$^{-1}$, $|\mathbf u|=7.4$~m~s$^{-1}$, which is reasonably close to the surface values in Fig.~\ref{fig:wind_speed}. 

\begin{figure}[h]
	\begin{center}
		{\includegraphics[width=5.0in]{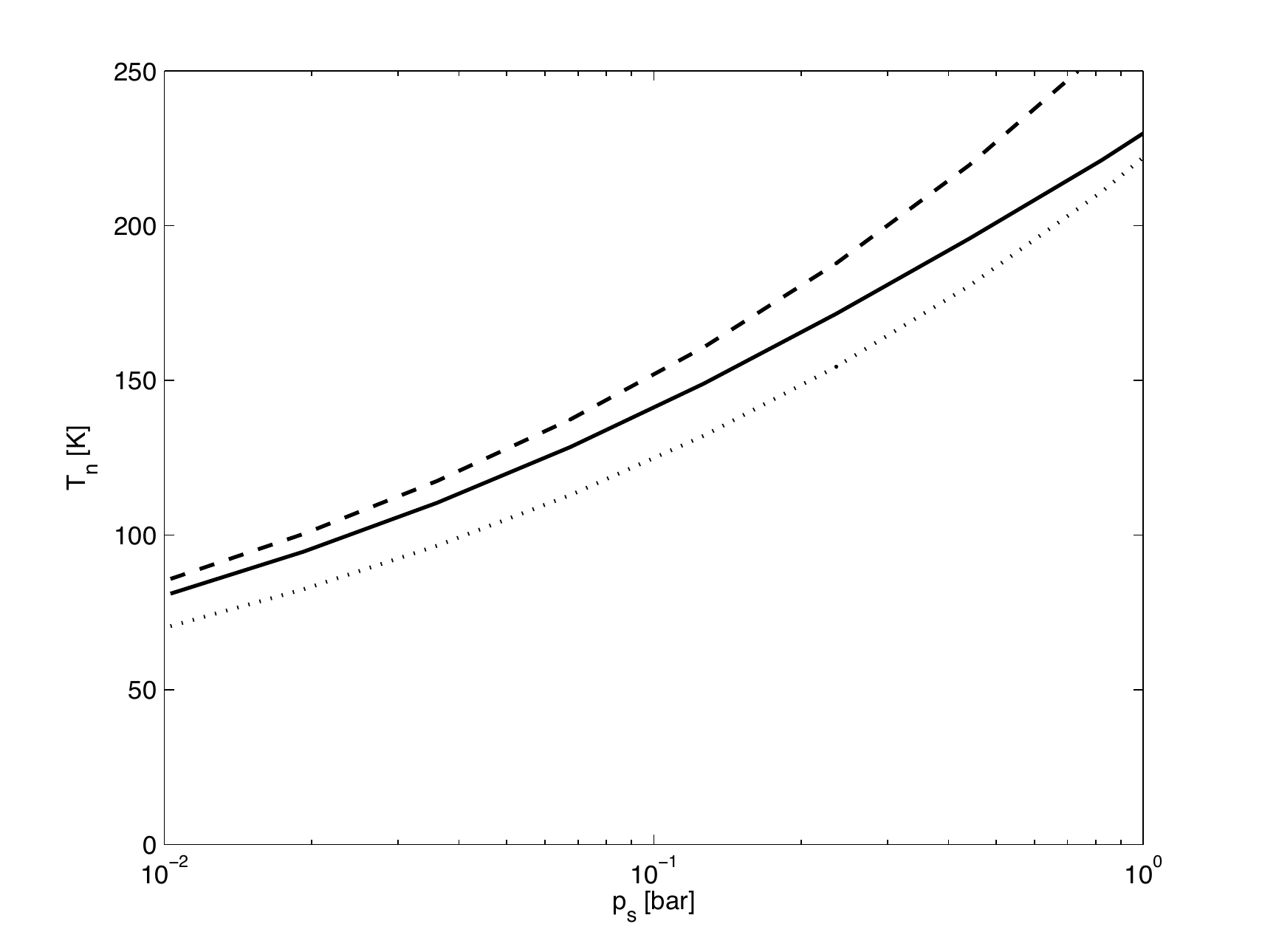}}
	\end{center}
	\caption{Mean nightside temperature vs. surface pressure in the gray GCM simulations (solid line), predicted by the radiative flux-only analysis (dotted line) and predicted by the radiative + sensible heat flux analysis (dashed line).}
	\label{fig:model_vs_theory}
\end{figure}

Fig.~\ref{fig:model_vs_theory} shows the nightside surface temperature predicted by (\ref{eq:night_thin2}) given this result vs. gray gas results from the GCM for a range of surface pressures. As can be seen, the correspondence is close over most of the range of pressures studied, with (\ref{eq:night_thin2}) slightly overpredicting the GCM value of $T_n$. The `thin radiator temperature' (\ref{eq:night_thin}) underpredicts the GCM results by a larger margin. The divergence around 1~bar can be explained by the fact that the atmosphere becomes optically thick at that pressure. A similar approach to scaling for the optically thick case should also be possible, and will be addressed in future work.

\subsection{Multiband 3D GCM simulations}

Having developed a comprehensive understanding of the key features of interhemispheric heat transport in the gray gas simulations, we now address the full atmospheric collapse problem in the GCM with correlated-$k$ radiative transfer. Figure~\ref{fig:Tprof_corrk} shows surface and atmospheric temperatures in 0.1~bar GCM simulations with the same planetary parameters as for Fig.~\ref{fig:Tslice} but using correlated-$k$ radiative transfer and assuming \ce{CO2} (black lines) and \ce{CO} (red lines) composition. 
The dotted lines on the left indicate condensation temperatures at 0.1~bar. As can be seen, the nightside is around 70~K colder for \ce{CO} than for \ce{CO2}. Nonetheless, the greater volatility of \ce{CO} means that it remains stable, whereas in the \ce{CO2} simulation the nightside temperature dips below the condensation temperature.

The inefficient heat transport of the \ce{CO} atmosphere can be understood by comparing the infrared absorption spectra of \ce{CO2} and \ce{CO} (Figure~\ref{fig:LBL}). As can be seen, the absorption bands of \ce{CO2} are wide and fall near the peak of the Planck function at both temperatures, while those of \ce{CO} are thinner and far from the Planck function peak, particularly at low temperatures. The ultimate reason for this is molecular structure: \ce{CO} is diatomic, with a permanent dipole, and so the infrared absorption spectrum consists of a single weak rotation band at low wave numbers and the fundamental vibration-rotation band centered at 2143.27~cm$^{-1}$ \citep{Goody1989}. \ce{CO2}, in contrast, has no rotation band due to the lack of a permanent dipole but intense vibration-rotation bands centered on 667~cm$^{-1}$ (15~$\mu$m) and 2325~cm$^{-1}$ (4.3~$\mu$m) due to the $\nu_2$ and $\nu_3$ fundamental modes, with complications due to Fermi resonances, Coriolis interactions and increased occupancy of higher vibrational levels at moderate temperatures. As demonstrated by the analysis in the preceding sections, the nightside temperature depends critically on the atmospheric opacity, and hence weakly absorbing gases cause extremely high surface temperature contrasts. In future observations of hot rocky planets, it should be possible to utilize this effect to constrain atmospheric properties. 

The remarkably high \emph{atmospheric} temperatures in the \ce{CO} case (Figure~\ref{fig:LBL}; right) appear paradoxical, but they can also be explained by the radiative properties of the gas. \ce{CO} is inefficient enough at radiating energy either to space or the surface that almost the entire atmosphere thermally equilibrates with the part of the surface where sensible heat fluxes peak: the dayside near the substellar point. Clearly, if more radiatively active gases or aerosols were present in a \ce{CO}-dominated atmosphere even in trace amounts, the behavior of the system could change considerably. Because of the subtlety of this problem, we leave the calculation of collapse pressure for \ce{CO} and other radiatively inactive, volatile gases such as \ce{N2} to future work. 
 
 With the atmospheric composition restricted to \ce{CO2} only, condensation was included, and GCM simulations were performed on a 14$\times$12 grid in the space of stellar flux and initial atmospheric pressure, for a planet with Earth's mass and radius. Figure~\ref{fig:condense_map} shows the results. The contour values show minimum surface temperature $T_{min}$.  In the white regions of the plot, $T_{min}<T_{cond}$, where $T_{cond}$ is the condensation temperature, and the atmosphere has begun to collapse on the surface. 

As can be seen, the onset of atmospheric collapse is a strong function of surface pressure, with collapse occurring for any stellar flux given $p_s<0.1$~bar. Collapse also occurs at all surface pressures for the lowest stellar flux studied (0.2~$F_0$ or $273.2$~W~m$^{-2}$). These fluxes are below even the values for the exoplanet candidate\footnote{Although believed to be a real planet by  virtually all observational groups for 7~years after its discovery \citep{Udry2007}, GJ581d is now disputed. See the discussion in \cite{Baluev2013} and \cite{Robertson2014} for details.} GJ581d, which we have shown can support stable \ce{CO2} atmospheres above pressures of around 10~bar \citep{Wordsworth2011}. The approximate Kombayshi-Ingersoll limit for \ce{CO2} \citep{Pierrehumbert2011b} is indicated by the vertical green dashed line.

The prediction of \ce{CO2} collapse at around 0.1~bar contrasts with the results of \cite{Joshi1997}, who found that  \ce{CO2} atmospheres with pressures as low as 30~mbar would be stable against collapse. There are two reasons for this discrepancy. First, as discussed earlier, in Joshi~et~al. predict somewhat higher nightside temperatures than found here even when we use the same gray gas radiative transfer as they did. Second, models with gray gas radiative transfer significantly overestimate coupling between the atmosphere and nightside surface\footnote{This effect was also pointed out in \cite{Leconte2013}.}. Particularly at pressures less than 1~bar, windows in the absorption spectrum (e.g., at wavenumbers less than 500~cm$^{-1}$ for \ce{CO2}) allow the surface to emit radiation directly to space, greatly increasing cooling rates.

It is also interesting to compare the results given here with the analytical predictions of \cite{Heng2012}. Essentially the entire range spanned by Fig.~\ref{fig:condense_map} is in their predicted `stable atmosphere' region  (see their Fig.~3), indicating that their $t_{adv}<t_{rad}$ criterion does not represent a sufficient condition for atmospheric stability independent of atmospheric pressure. Conversely,  $t_{adv}>t_{rad}$  does not appear to be sufficient for instability in all cases, because in simulations assuming a 10~M$_\Earth$ planet, stable atmospheric solutions at high \ce{CO2} pressures were also found [Fig.~\ref{fig:condense_map} (bottom)]. In the highly irradiated regime where the flow Mach number can approach the sound speed, however, their approach may still be applicable. Applying 3D circulation models with real gas radiative transfer to this problem will be an interesting topic for future research. 

Finally, using the data shown in Figure~\ref{fig:condense_map}, empirical formulae of the critical collapse pressure $p_{crit}$ vs. stellar flux were created according to the expression
\begin{equation}
log_{10}\left[ \frac{p_{crit}}{1~\mbox{Pa}} \right] = c_1 F_s^3 + c_2 F_s^2 + c_3 F_s + c_4 \label{eq:pcollapse_empiric}
\end{equation}
with the constants $c_i$ derived by a least-squares fit. The results are shown in Figure~\ref{fig:condense_map}, with the coefficients given in Table~\ref{tab:coeffs}. 

The atmosphere collapses at higher pressures in general on higher mass planets for several reasons. First, the increase in gravity means a lower column mass for a given pressure. This means the total atmospheric opacity is also lower, and hence the nightside temperature decreases. In addition, higher planetary radii mean the advection time $t_{adv} = r_p\slash |\mathbf u|$ should be smaller if $|\mathbf u|$ remains constant. However, in the WTG regime the scaling of $\tilde U$ with $\tilde L$ and hence $r_p$ suggests that this effect should be less important. 

The decrease in the critical collapse pressure with stellar flux can be understood by reference to (\ref{eq:Tn_simplest}). Increasing stellar flux implies a greater amount of heat is transported to the nightside, and hence a lower total pressure can be achieved before collapse occurs. Because at fixed stellar luminosity we increase stellar flux by moving the planet inwards, the effects of rotation on the circulation must eventually become important. However, its importance appears to be secondary for the range of cases studied here.

\section{Discussion}\label{sec:discuss}

Atmospheric collapse is likely to be of fundamental importance for tidally locked rocky planets. The analysis in the preceding section has demonstrated that despite the fact that the collapse problem involves coupling between several complex processes (radiative transfer, the large-scale circulation and boundary layer turbulence), its key features can be understood using an entirely analytical approach.

How robust are these results? Clearly,  modifying effects such as ocean heat transport or intense tidal heating of the planetary interior could result in more stable atmospheres than found here. The presence of radiatively active trace gases would also render an atmosphere more stable than the single-component cases studied here, by increasing the radiative coupling between the atmosphere and night side of the planet. A similar role could be played by aerosols released by volcanism or by surface lifting of dust. Clouds can also potentially contribute to greenhouse warming on the nightside, as we found previously for \ce{CO2} in \cite{Wordsworth2011}, although thermodynamics disfavours nightside \ce{H2O} cloud formation for Earth-like tidally locked planets \citep{Yang2013}. Further modeling will be required to assess the potential importance of these effects. Most processes will lead to a reduction in the collapse pressure, though, so the results in Table~\ref{tab:coeffs} can be regarded as a conservative upper limit.

Another point worth discussing is the representation of the boundary layer physics in the strongly stratified regime. Evidence from terrestrial observations of nocturnal and polar planetary boundary layers indicates that turbulent motion does not cease completely when $Ri$ is greater than the `critical' value of $0.2$  \citep{Strang2001,Monti2002,Galperin2007}. This effect, which can be understood in the context of gravity wave turbulence, is not properly accounted for in the MYG boundary layer scheme.  Could it significantly modify the results described here?

\begin{figure}[h]
	\begin{center}
		{\includegraphics[width=6.5in]{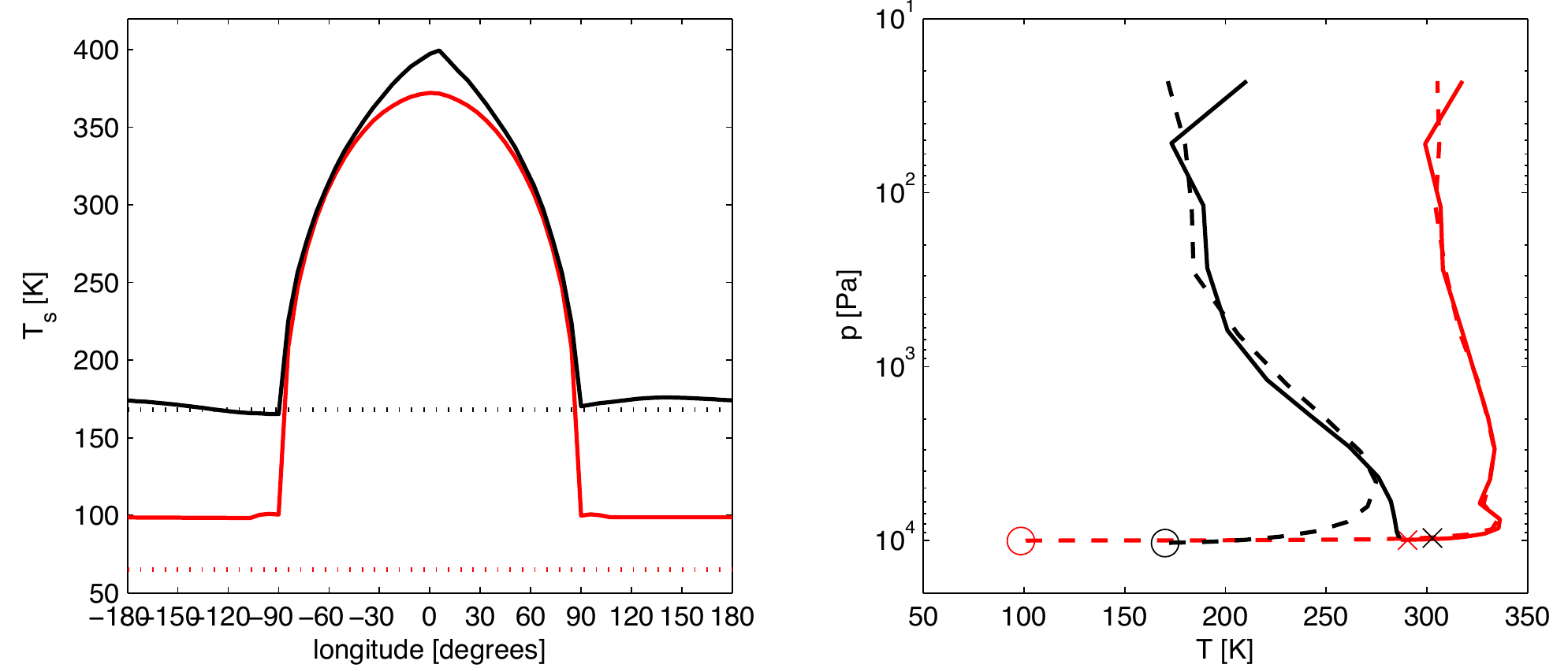}}
	\end{center}
	\caption{(left) Time averaged equatorial surface temperatures at zero latitude in the correlated-$k$ simulations for \ce{CO2} (black) and \ce{CO} (red). The dotted lines indicate the condensation temperature at 0.1~bar for each species. (right) Hemisphere and time-averaged day and nightside temperatures vs. pressure in the same simulations. Crosses and circles indicate day- and nightside surface temperatures, respectively. }
	\label{fig:Tprof_corrk}
\end{figure}

\begin{figure}[h]
	\begin{center}
		{\includegraphics[width=4.0in]{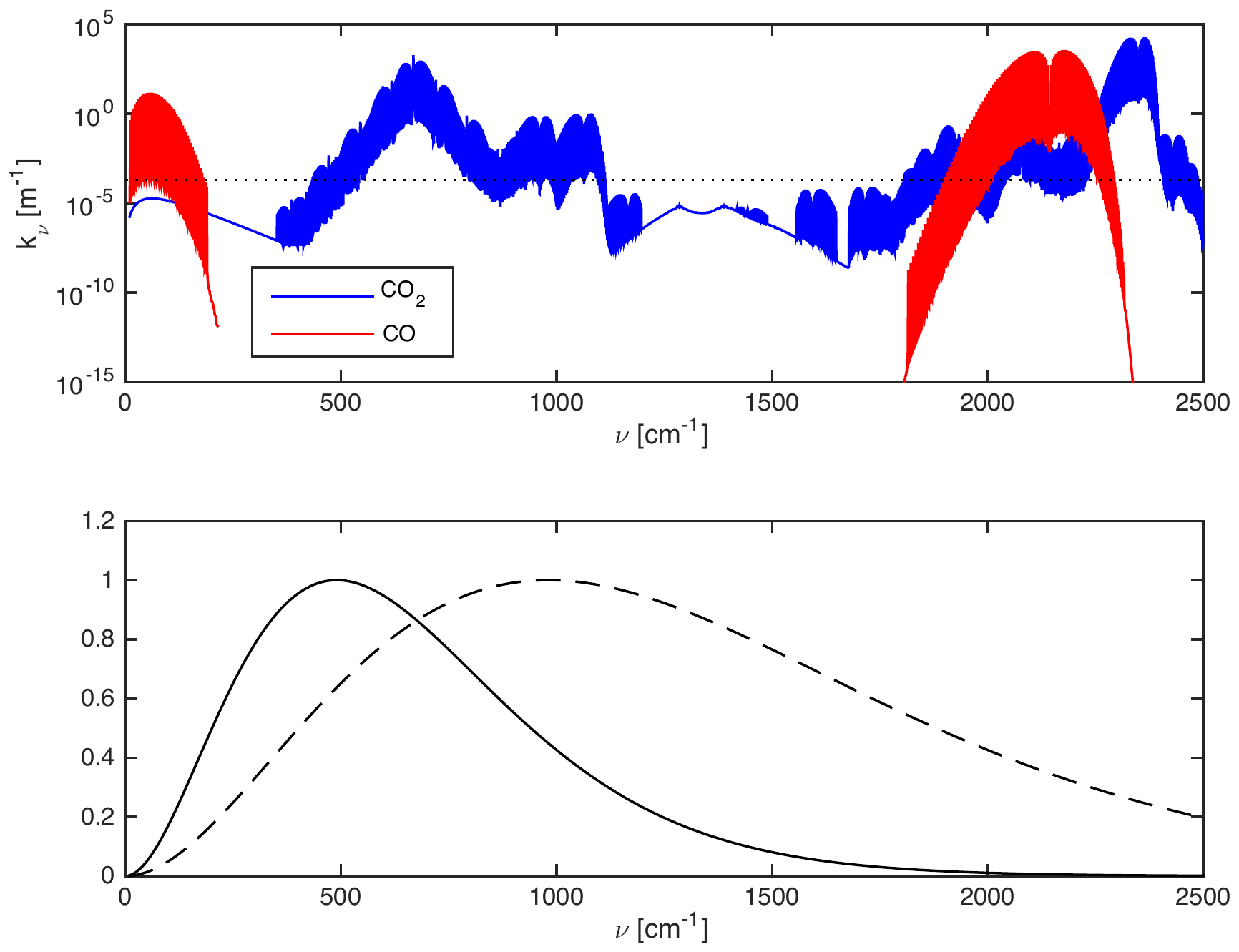}}
	\end{center}
	\caption{(top) Infrared absorptivity $k_\nu$ of \ce{CO} and \ce{CO2} at 400~K and 0.1~bar. The black line indicates $\tau=H k_\nu = 1$ for $H = 5$~km. \ce{CO2} collision-induced absorption is included for completeness, but its effects are weak at 0.1~bar \citep{Wordsworth2010,Gruszka1997,Baranov2004}. (bottom) Normalized Planck function at 250~K (solid line) and 500~K (dashed line). }
	\label{fig:LBL}
\end{figure}

\begin{figure}[h]
	\begin{center}
		{\includegraphics[width=4.0in]{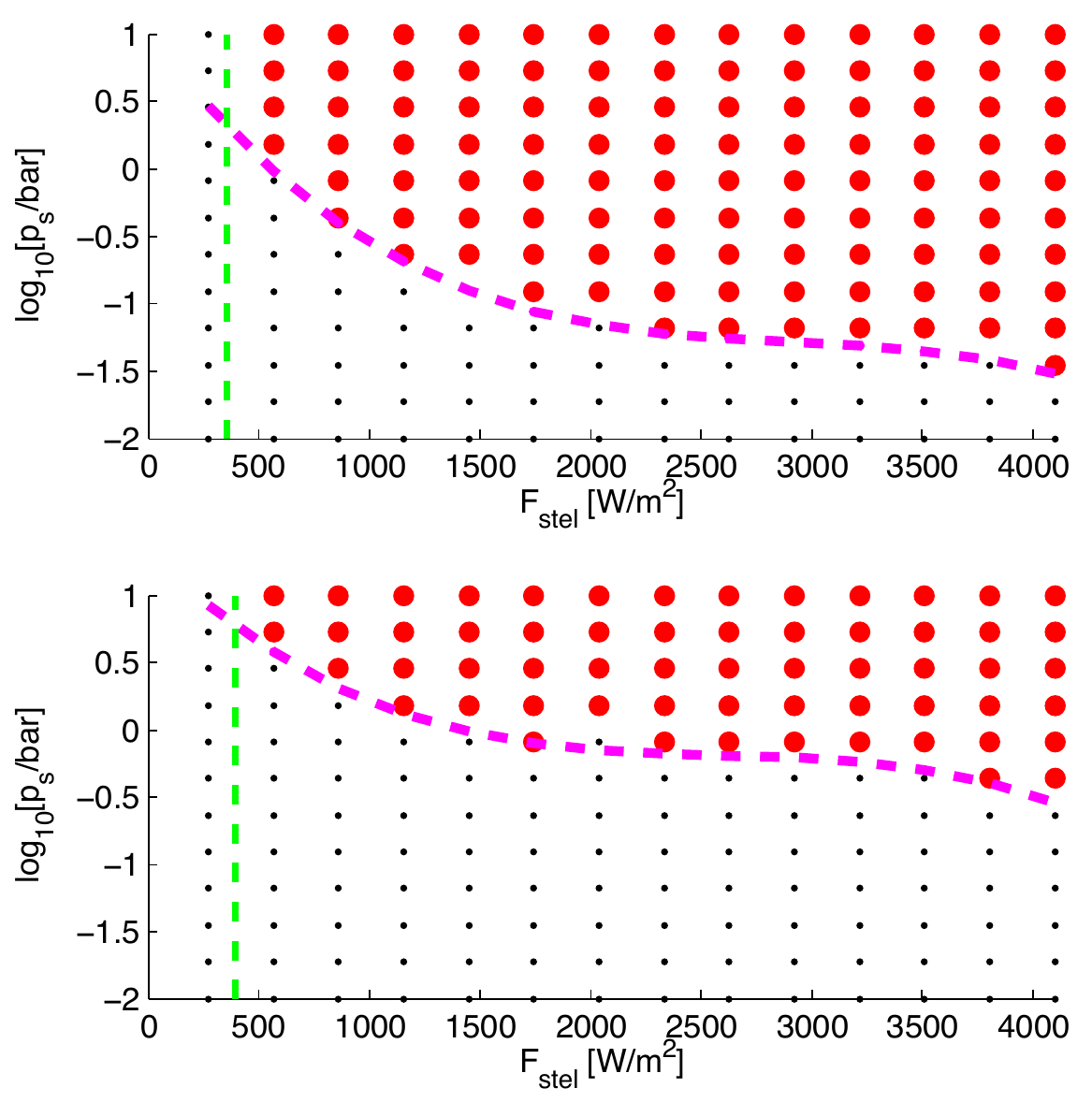}}
	\end{center}
	\caption{Stability diagram for the \ce{CO2} simulations with correlated-$k$ radiative transfer for a 1~$M_E$ planet (top) and a 10~$M_E$ planet (bottom). Large blue dots indicate simulations where the atmosphere remained stable. 
	The green dashed line indicates the approximate K-I limit for \ce{CO2}, while the magenta dashed line indicates the empirical collapse curve (\ref{eq:pcollapse_empiric}) derived from the data.}
	\label{fig:condense_map}
\end{figure}

\cite{Sukoriansky2005} recently developed a spectral model of stratified turbulent flows that accounts for gravity wave turbulence in the high $Ri$ regime, using a renormalization group approach that progressively replaces the full nonlinear fluid equations at increasing scales by a quasi-Gaussian Langevin equation. They showed that as $Ri$ increases, horizontal eddy viscosities and diffusivities increase, enhancing horizontal mixing. However, vertical eddy viscosity drops to a low (constant) value, while vertical diffusivity declines to zero [Fig.~8 in \cite{Sukoriansky2005}]. At sufficiently high stable stratification, the sensible heat flux should be negligible, even while horizontal mixing remains high. In the 0.1~bar \ce{CO2} simulation in Fig.~\ref{fig:Tprof_corrk} of this paper, the bulk Richardson number $Ri_B$ at the surface is around 1, implying a small addition to the sensible heat flux due to gravity wave turbulence. In future it will be interesting to incorporate this effect into the GCM boundary layer scheme. It is likely to be more significant for \ce{CO} and other highly volatile, radiatively inactive gases than for gases like \ce{CO2}.

For potentially habitable planets, this work has several broad implications. Clearly, atmospheric collapse is a vital effect that may be a key driver of climate in some cases. It has been shown that interhemispheric heat transport and hence global climate depends strongly on both total atmospheric pressure and composition --- a fact that is often neglected in GCM studies of habitability around M-stars, which currently tend to focus on Earth-like planets only. Here, only \ce{CO2} and \ce{CO} atmospheres have been studied, to allow a focus on physical processes, but collapse of other volatiles on a planet's nightside, particularly \ce{H2O} [e.g., \cite{Leconte2013,Menou2013}], may be significant and linked to the overall oxidation rate of the planet \citep{Wordsworth2014}.  Future work needs to focus on the efficiency of delivery/loss mechanisms for highly volatile gases (\ce{H2}, \ce{N2}, \ce{Ar}, \ce{He}) on low-mass planets, and coupling between the dynamical processes studied here and atmospheric chemistry, including the overall rate of oxidation via photolysis and hydrogen escape for planets with surface liquid water. 

Finally, this study has important potential implications for future observations of rocky exoplanets (habitable or not) by ground-based facilities or spacecraft such as JWST. Most simply, if a given molecule is detected in the transit spectrum of a tidally locked planet, knowledge of that molecule's critical collapse pressure will allow constraints to be placed on the atmospheric composition. Similarly, a rigorous understanding of rocky planet interhemispheric heat transport is necessary for future interpretation of future broadband transit and phase curve data. Finally, as discussed in \cite{Selsis2011}, if a planet's variation spectrum can also be retrieved, powerful constraints on atmospheric pressure and composition are possible. Future work will generalize the results presented here to a variety of atmospheric compositions and investigate these issues in greater detail.

\acknowledgments

This research was partially supported by the National Science Foundation and NASA's VPL program. This article benefited from discussions with many researchers, including Peter Read, Bob Haberle, Zhiming Kuang and Remco de Kok. The computations in this paper were run on the Odyssey cluster supported by the FAS Division of Science, Research Computing Group at Harvard University.

\bibliography{allrefs}

\begin{thebibliography}{}

\bibitem[Baluev, 2013]{Baluev2013}
Baluev, R.~V. (2013).
\newblock The impact of red noise in radial velocity planet searches: only
  three planets orbiting gj 581?
\newblock {\em Monthly Notices of the Royal Astronomical Society},
  429(3):2052--2068.

\bibitem[{Baranov} et~al., 2004]{Baranov2004}
{Baranov}, Y.~I., {Lafferty}, W.~J., and {Fraser}, G.~T. (2004).
\newblock {Infrared spectrum of the continuum and dimer absorption in the
  vicinity of the O2 vibrational fundamental in O2/CO2 mixtures}.
\newblock {\em J. Mol. Spectrosc.}, 228:432--440.

\bibitem[{Bean} et~al., 2010]{Bean2010}
{Bean}, J.~L., {Kempton}, E., and {Homeier}, D. (2010).
\newblock {A ground-based transmission spectrum of the super-Earth exoplanet GJ
  1214b}.
\newblock {\em Nature}, 468:669--672.

\bibitem[Blackadar, 1962]{Blackadar1962}
Blackadar, A.~K. (1962).
\newblock The vertical distribution of wind and turbulent exchange in a neutral
  atmosphere.
\newblock {\em Journal of Geophysical Research}, 67(8):3095--3102.

\bibitem[Castan and Menou, 2011]{Castan2011}
Castan, T. and Menou, K. (2011).
\newblock Atmospheres of hot super-earths.
\newblock {\em The Astrophysical Journal Letters}, 743(2):L36.

\bibitem[Cerni and Parish, 1984]{Cerni1984}
Cerni, T.~A. and Parish, T.~R. (1984).
\newblock A radiative model of the stable nocturnal boundary layer with
  application to the polar night.
\newblock {\em Journal of Climate and Applied Meteorology}, 23(11):1563--1572.

\bibitem[Charbonneau et~al., 2009]{Charbonneau2009}
Charbonneau, D., Berta, Z.~K., Irwin, J., Burke, C.~J., Nutzman, P., Buchhave,
  L.~A., Lovis, C., Bonfils, X., Latham, D.~W., Udry, S., et~al. (2009).
\newblock A super-earth transiting a nearby low-mass star.
\newblock {\em Nature}, 462(7275):891--894.

\bibitem[Cohen et~al., 2014]{Cohen2014}
Cohen, O., Drake, J., Glocer, A., Garraffo, C., Poppenhaeger, K., Bell, J.,
  Ridley, A., and Gombosi, T. (2014).
\newblock Magnetospheric structure and atmospheric joule heating of habitable
  planets orbiting m-dwarf stars.
\newblock {\em arXiv preprint arXiv:1405.7707}.

\bibitem[Croll et~al., 2011]{Croll2011}
Croll, B., Albert, L., Jayawardhana, R., Kempton, E. M.-R., Fortney, J.~J.,
  Murray, N., and Neilson, H. (2011).
\newblock Broadband transmission spectroscopy of the super-earth gj 1214b
  suggests a low mean molecular weight atmosphere.
\newblock {\em The Astrophysical Journal}, 736(2):78.

\bibitem[Demory et~al., 2012]{Demory2012}
Demory, B.-O., Gillon, M., Seager, S., Benneke, B., Deming, D., and Jackson, B.
  (2012).
\newblock Detection of thermal emission from a super-earth.
\newblock {\em The Astrophysical Journal Letters}, 751(2):L28.

\bibitem[Downs et~al., 1975]{Downs1975}
Downs, G., Reichley, P., and Green, R. (1975).
\newblock Radar measurements of martian topography and surface properties: The
  1971 and 1973 oppositions.
\newblock {\em Icarus}, 26(3):273--312.

\bibitem[{Edson} et~al., 2012]{Edson2012}
{Edson}, A.~R., {Kasting}, J.~F., {Pollard}, D., {Lee}, S., and {Bannon}, P.~R.
  (2012).
\newblock {The Carbonate-Silicate Cycle and CO2/Climate Feedbacks on Tidally
  Locked Terrestrial Planets}.
\newblock {\em Astrobiology}, 12:562--571.

\bibitem[{Forget} et~al., 1999]{Forget1999}
{Forget}, F., {Hourdin}, F., {Fournier}, R., {Hourdin}, C., {Talagrand}, O.,
  {Collins}, M., {Lewis}, S.~R., {Read}, P.~L., and {Huot}, J. (1999).
\newblock {Improved general circulation models of the Martian atmosphere from
  the surface to above 80 km}.
\newblock {\em Journal of Geophysical Research}, 104:24155--24176.

\bibitem[Fraine et~al., 2014]{Fraine2014}
Fraine, J., Deming, D., Benneke, B., Knutson, H., Jord{\'a}n, A., Espinoza, N.,
  Madhusudhan, N., Wilkins, A., and Todorov, K. (2014).
\newblock Water vapour absorption in the clear atmosphere of a neptune-sized
  exoplanet.
\newblock {\em Nature}, 513(7519):526--529.

\bibitem[Frommhold, 2006]{Frommhold2006}
Frommhold, L. (2006).
\newblock {\em Collision-induced absorption in gases}, volume~2.
\newblock Cambridge University Press.

\bibitem[{Galperin} et~al., 1988]{Galperin1988}
{Galperin}, B., {Kantha}, L.~H., {Hassid}, S., and {Rosati}, A. (1988).
\newblock {A Quasi-equilibrium Turbulent Energy Model for Geophysical Flows.}
\newblock {\em Journal of Atmospheric Sciences}, 45:55--62.

\bibitem[Galperin et~al., 2007]{Galperin2007}
Galperin, B., Sukoriansky, S., and Anderson, P.~S. (2007).
\newblock On the critical richardson number in stably stratified turbulence.
\newblock {\em Atmospheric Science Letters}, 8(3):65--69.

\bibitem[Garratt, 1994]{Garratt1994}
Garratt, J.~R. (1994).
\newblock {\em The atmospheric boundary layer}.
\newblock Cambridge University Press.

\bibitem[Goody and Yung, 1989]{Goody1989}
Goody, R.~M. and Yung, Y.~L. (1989).
\newblock Atmospheric radiation: theoretical basis.
\newblock {\em Atmospheric radiation: theoretical basis, 2nd ed., by Richard M.
  Goody and YL Yung. New York, NY: Oxford University Press, 1989}, 1.

\bibitem[{Gruszka} and {Borysow}, 1997]{Gruszka1997}
{Gruszka}, M. and {Borysow}, A. (1997).
\newblock {Roto-Translational Collision-Induced Absorption of CO2 for the
  Atmosphere of Venus at Frequencies from 0 to 250 cm\^{}-1, at Temperatures
  from 200 to 800 K}.
\newblock {\em Icarus}, 129:172--177.

\bibitem[{Gruszka} and {Borysow}, 1998]{Gruszka1998}
{Gruszka}, M. and {Borysow}, A. (1998).
\newblock {Computer simulation of the far infrared collision induced absorption
  spectra of gaseous CO2}.
\newblock {\em Molecular Physics}, 93:1007--1016.

\bibitem[Haberle et~al., 1993]{Haberle1993}
Haberle, R.~M., Houben, H.~C., Hertenstein, R., and Herdtle, T. (1993).
\newblock A boundary-layer model for mars: Comparison with viking lander and
  entry data.
\newblock {\em Journal of the atmospheric sciences}, 50(11):1544--1559.

\bibitem[Head et~al., 1985]{Head1985}
Head, J.~W., Peterfreund, A.~R., Garvin, J.~B., and Zisk, S.~H. (1985).
\newblock Surface characteristics of venus derived from pioneer venus
  altimetry, roughness, and reflectivity measurements.
\newblock {\em Journal of Geophysical Research: Solid Earth (1978--2012)},
  90(B8):6873--6885.

\bibitem[Heng and Kopparla, 2012]{Heng2012}
Heng, K. and Kopparla, P. (2012).
\newblock On the stability of super-earth atmospheres.
\newblock {\em The Astrophysical Journal}, 754(1):60.

\bibitem[Heng and Vogt, 2011]{Heng2011}
Heng, K. and Vogt, S.~S. (2011).
\newblock Gliese 581g as a scaled-up version of earth: atmospheric circulation
  simulations.
\newblock {\em Monthly Notices of the Royal Astronomical Society},
  415(3):2145--2157.

\bibitem[Ingersoll et~al., 1985]{Ingersoll1985}
Ingersoll, A.~P., Summers, M.~E., and Schlipf, S.~G. (1985).
\newblock Supersonic meteorology of io: Sublimation-driven flow of so2.
\newblock {\em Icarus}, 64(3):375--390.

\bibitem[{Joshi}, 2003]{Joshi2003}
{Joshi}, M. (2003).
\newblock {Climate Model Studies of Synchronously Rotating Planets}.
\newblock {\em Astrobiology}, 3:415--427.

\bibitem[Joshi et~al., 1995]{Joshi1995}
Joshi, M., Lewis, S., Read, P., and Catling, D. (1995).
\newblock Western boundary currents in the martian atmosphere: Numerical
  simulations and observational evidence.
\newblock {\em Journal of Geophysical Research: Planets (1991--2012)},
  100(E3):5485--5500.

\bibitem[{Joshi} et~al., 1997]{Joshi1997}
{Joshi}, M.~M., {Haberle}, R.~M., and {Reynolds}, R.~T. (1997).
\newblock {Simulations of the Atmospheres of Synchronously Rotating Terrestrial
  Planets Orbiting M Dwarfs: Conditions for Atmospheric Collapse and the
  Implications for Habitability}.
\newblock {\em Icarus}, 129:450--465.

\bibitem[Kaspi and Showman, 2014]{Kaspi2014}
Kaspi, Y. and Showman, A.~P. (2014).
\newblock Atmospheric dynamics of terrestrial exoplanets over a wide range of
  orbital and atmospheric parameters.
\newblock {\em arXiv preprint arXiv:1407.6349}.

\bibitem[{Kasting} et~al., 1993]{Kasting1993}
{Kasting}, J.~F., {Whitmire}, D.~P., and {Reynolds}, R.~T. (1993).
\newblock {Habitable Zones around Main Sequence Stars}.
\newblock {\em Icarus}, 101:108--128.

\bibitem[Khodachenko et~al., 2007]{Khodachenko2007}
Khodachenko, M.~L., Ribas, I., Lammer, H., Grie{\ss}meier, J.-M., Leitner, M.,
  Selsis, F., Eiroa, C., Hanslmeier, A., Biernat, H.~K., Farrugia, C.~J.,
  et~al. (2007).
\newblock Coronal mass ejection (cme) activity of low mass m stars as an
  important factor for the habitability of terrestrial exoplanets. i. cme
  impact on expected magnetospheres of earth-like exoplanets in close-in
  habitable zones.
\newblock {\em Astrobiology}, 7(1):167--184.

\bibitem[{Kreidberg} et~al., 2013]{Kreidberg2013}
{Kreidberg}, L., {Bean}, J., {D{\'e}sert}, J., {Seager}, S., {Deming}, D.,
  {Benneke}, B., {Berta}, Z.~K., {Stevenson}, K.~B., and {Homeier}, D. (2013).
\newblock {Transmission Spectroscopy of the Super-Earth GJ 1214b Using HST/WFC3
  in Spatial Scan Mode}.
\newblock In {\em American Astronomical Society Meeting Abstracts}, volume 221
  of {\em American Astronomical Society Meeting Abstracts}, page 224.03.

\bibitem[{Lammer} et~al., 2007]{Lammer2007}
{Lammer}, H., {Lichtenegger}, H.~I.~M., {Kulikov}, Y.~N., {Grie{\ss}meier}, J.,
  {Terada}, N., {Erkaev}, N.~V., {Biernat}, H.~K., {Khodachenko}, M.~L.,
  {Ribas}, I., {Penz}, T., and {Selsis}, F. (2007).
\newblock {Coronal Mass Ejection (CME) Activity of Low Mass M Stars as An
  Important Factor for The Habitability of Terrestrial Exoplanets. II.
  CME-Induced Ion Pick Up of Earth-like Exoplanets in Close-In Habitable
  Zones}.
\newblock {\em Astrobiology}, 7:185--207.

\bibitem[Leconte et~al., 2013]{Leconte2013}
Leconte, J., Forget, F., Charnay, B., Wordsworth, R., Selsis, F., and Millour,
  E. (2013).
\newblock 3d climate modeling of close-in land planets: Circulation patterns,
  climate moist bistability and habitability.
\newblock {\em arXiv preprint arXiv:1303.7079}.

\bibitem[Lide, 2000]{CRC2000}
Lide, D.~P., editor (2000).
\newblock {\em CRC Handbook of Chemistry and Physics}.
\newblock CRC PRESS, 81 edition.

\bibitem[Linsky et~al., 2013]{Linsky2013}
Linsky, J.~L., France, K., and Ayres, T. (2013).
\newblock Computing intrinsic {LY}$\alpha$ fluxes of {F5 V} to {M5 V} stars.
\newblock {\em The Astrophysical Journal}, 766(2):69.

\bibitem[{Mayor} et~al., 2009]{Mayor2009}
{Mayor}, M., {Bonfils}, X., {Forveille}, T., {Delfosse}, X., {Udry}, S.,
  {Bertaux}, J., {Beust}, H., {Bouchy}, F., {Lovis}, C., {Pepe}, F., {Perrier},
  C., {Queloz}, D., and {Santos}, N.~C. (2009).
\newblock {The HARPS search for southern extra-solar planets. XVIII. An
  Earth-mass planet in the GJ 581 planetary system}.
\newblock {\em Astronomy and Astrophysics}, 507:487--494.

\bibitem[{Mellor} and {Yamada}, 1982]{Mellor1982}
{Mellor}, G.~L. and {Yamada}, T. (1982).
\newblock {Development of a Turbulence Closure Model for Geophysical Fluid
  Problems}.
\newblock {\em Reviews of Geophysics}, 20:851--875.

\bibitem[Menou, 2013]{Menou2013}
Menou, K. (2013).
\newblock Water-trapped worlds.
\newblock {\em The Astrophysical Journal}, 774(1):51.

\bibitem[Merlis and Schneider, 2010]{Merlis2010}
Merlis, T.~M. and Schneider, T. (2010).
\newblock Atmospheric dynamics of earth-like tidally locked aquaplanets.
\newblock {\em Journal of Advances in Modeling Earth Systems}, 2(4).

\bibitem[Miguel et~al., 2011]{Miguel2011}
Miguel, Y., Kaltenegger, L., Fegley, B., and Schaefer, L. (2011).
\newblock Compositions of hot super-earth atmospheres: exploring kepler
  candidates.
\newblock {\em The Astrophysical Journal Letters}, 742(2):L19.

\bibitem[Monti et~al., 2002]{Monti2002}
Monti, P., Fernando, H., Princevac, M., Chan, W., Kowalewski, T., and Pardyjak,
  E. (2002).
\newblock Observations of flow and turbulence in the nocturnal boundary layer
  over a slope.
\newblock {\em Journal of the Atmospheric Sciences}, 59(17):2513--2534.

\bibitem[Pepe et~al., 2011]{Pepe2011}
Pepe, F., Lovis, C., Segransan, D., Benz, W., Bouchy, F., Dumusque, X., Mayor,
  M., Queloz, D., Santos, N., and Udry, S. (2011).
\newblock The harps search for earth-like planets in the habitable zone:
  I--very low-mass planets around hd20794, hd85512 and hd192310.
\newblock {\em arXiv preprint arXiv:1108.3447}.

\bibitem[Perez-Becker and Showman, 2013]{Perez2013}
Perez-Becker, D. and Showman, A.~P. (2013).
\newblock Atmospheric heat redistribution on hot jupiters.
\newblock {\em The Astrophysical Journal}, 776(2):134.

\bibitem[{Pettersen} and {Coleman}, 1981]{Pettersen1981}
{Pettersen}, B.~R. and {Coleman}, L.~A. (1981).
\newblock {Chromospheric lines in red dwarf flare stars. I - AD Leonis and GX
  Andromedae}.
\newblock {\em \apj}, 251:571--582.

\bibitem[Pierrehumbert, 2011a]{Pierrehumbert2011BOOK}
Pierrehumbert, R. (2011a).
\newblock {\em Principles of Planetary Climate}.
\newblock Cambridge University Press.

\bibitem[Pierrehumbert, 2011b]{Pierrehumbert2011b}
Pierrehumbert, R.~T. (2011b).
\newblock {A palette of climates for {G}liese 581g}.
\newblock {\em The Astrophysical Journal Letters}, 726(1):L8.

\bibitem[Reid et~al., 2000]{Reid2000}
Reid, N., Reid, I.~N., Reid, N., and Hawley, S. (2000).
\newblock {\em New light on dark stars}.
\newblock Springer.

\bibitem[Reiners et~al., 2009]{Reiners2009}
Reiners, A., Basri, G., and Browning, M. (2009).
\newblock Evidence for magnetic flux saturation in rapidly rotating m stars.
\newblock {\em The Astrophysical Journal}, 692(1):538.

\bibitem[Robertson et~al., 2014]{Robertson2014}
Robertson, P., Mahadevan, S., Endl, M., and Roy, A. (2014).
\newblock Stellar activity masquerading as planets in the habitable zone of the
  m dwarf gliese 581.
\newblock {\em Science}, 345(6195):440--444.

\bibitem[Rosenburg et~al., 2011]{Rosenburg2011}
Rosenburg, M., Aharonson, O., Head, J., Kreslavsky, M., Mazarico, E., Neumann,
  G.~A., Smith, D.~E., Torrence, M.~H., and Zuber, M.~T. (2011).
\newblock Global surface slopes and roughness of the moon from the lunar
  orbiter laser altimeter.
\newblock {\em Journal of Geophysical Research: Planets (1991--2012)}, 116(E2).

\bibitem[Rothman et~al., 2010]{Rothman2010}
Rothman, L., Gordon, I., Barber, R., Dothe, H., Gamache, R., Goldman, A.,
  Perevalov, V., Tashkun, S., and Tennyson, J. (2010).
\newblock Hitemp, the high-temperature molecular spectroscopic database.
\newblock {\em Journal of Quantitative Spectroscopy and Radiative Transfer},
  111(15):2139--2150.

\bibitem[{Rothman} et~al., 2009]{Rothman2009}
{Rothman}, L.~S., {Gordon}, I.~E., {Barbe}, A., {Benner}, D.~C., {Bernath},
  P.~F., {Birk}, M., {Boudon}, V., {Brown}, L.~R., {Campargue}, A., {Champion},
  J.-P., {Chance}, K., {Coudert}, L.~H., {Dana}, V., {Devi}, V.~M., {Fally},
  S., {Flaud}, J.-M., {Gamache}, R.~R., {Goldman}, A., {Jacquemart}, D.,
  {Kleiner}, I., {Lacome}, N., {Lafferty}, W.~J., {Mandin}, J.-Y., {Massie},
  S.~T., {Mikhailenko}, S.~N., {Miller}, C.~E., {Moazzen-Ahmadi}, N.,
  {Naumenko}, O.~V., {Nikitin}, A.~V., {Orphal}, J., {Perevalov}, V.~I.,
  {Perrin}, A., {Predoi-Cross}, A., {Rinsland}, C.~P., {Rotger}, M., {{\v
  S}ime{\v c}kov{\'a}}, M., {Smith}, M.~A.~H., {Sung}, K., {Tashkun}, S.~A.,
  {Tennyson}, J., {Toth}, R.~A., {Vandaele}, A.~C., and {Vander Auwera}, J.
  (2009).
\newblock {The HITRAN 2008 molecular spectroscopic database}.
\newblock {\em Journal of Quantitative Spectroscopy and Radiative Transfer},
  110:533--572.

\bibitem[Samuel et~al., 2014]{Samuel2014}
Samuel, B., Leconte, J., Rouan, D., Forget, F., L{\'e}ger, A., and Schneider,
  J. (2014).
\newblock Constraining physics of very hot super-earths with the james webb
  telescope. the case of corot-7b.
\newblock {\em arXiv preprint arXiv:1402.6637}.

\bibitem[Seager et~al., 2010]{Seager2010}
Seager, S., Dotson, R., et~al. (2010).
\newblock {\em Exoplanets}.
\newblock University of Arizona Press.

\bibitem[{Segura} et~al., 2003]{Segura2003}
{Segura}, A., {Krelove}, K., {Kasting}, J.~F., {Sommerlatt}, D., {Meadows}, V.,
  {Crisp}, D., {Cohen}, M., and {Mlawer}, E. (2003).
\newblock {Ozone Concentrations and Ultraviolet Fluxes on Earth-Like Planets
  Around Other Stars}.
\newblock {\em Astrobiology}, 3:689--708.

\bibitem[Selsis et~al., 2011]{Selsis2011}
Selsis, F., Wordsworth, R., and Forget, F. (2011).
\newblock Thermal phase curves of nontransiting terrestrial exoplanets: I.
  characterizing atmospheres.
\newblock {\em Astronomy \& Astrophysics}, 532.

\bibitem[{Shkolnik} et~al., 2009]{Shkolnik2009}
{Shkolnik}, E., {Liu}, M.~C., and {Reid}, I.~N. (2009).
\newblock {Identifying the Young Low-mass Stars within 25 pc. I. Spectroscopic
  Observations}.
\newblock {\em \apj}, 699:649--666.

\bibitem[Sobel et~al., 2001]{Sobel2001}
Sobel, A.~H., Nilsson, J., and Polvani, L.~M. (2001).
\newblock The weak temperature gradient approximation and balanced tropical
  moisture waves*.
\newblock {\em Journal of the atmospheric sciences}, 58(23):3650--3665.

\bibitem[{Sotin} et~al., 2007]{Sotin2007}
{Sotin}, C., {Grasset}, O., and {Mocquet}, A. (2007).
\newblock {Mass radius curve for extrasolar Earth-like planets and ocean
  planets}.
\newblock {\em Icarus}, 191:337--351.

\bibitem[Strang and Fernando, 2001]{Strang2001}
Strang, E. and Fernando, H. (2001).
\newblock Vertical mixing and transports through a stratified shear layer.
\newblock {\em Journal of physical oceanography}, 31(8):2026--2048.

\bibitem[Sukoriansky et~al., 2005]{Sukoriansky2005}
Sukoriansky, S., Galperin, B., and Staroselsky, I. (2005).
\newblock A quasinormal scale elimination model of turbulent flows with stable
  stratification.
\newblock {\em Physics of Fluids (1994-present)}, 17(8):085107.

\bibitem[{Tian}, 2009]{Tian2009}
{Tian}, F. (2009).
\newblock {Thermal Escape from Super Earth Atmospheres in the Habitable Zones
  of M Stars}.
\newblock {\em The Astrophysical Journal}, 703:905--909.

\bibitem[Tuomi et~al., 2012]{Tuomi2012}
Tuomi, M., Jones, H.~R., Jenkins, J.~S., Tinney, C.~G., Butler, R.~P., Vogt,
  S.~S., Barnes, J.~R., Wittenmyer, R.~A., O'Toole, S., Horner, J., et~al.
  (2012).
\newblock Signals embedded in the radial velocity noise. periodic variations in
  the tau ceti velocities.
\newblock {\em arXiv preprint arXiv:1212.4277}.

\bibitem[{Udry} et~al., 2007]{Udry2007}
{Udry}, S., {Bonfils}, X., {Delfosse}, X., {Forveille}, T., {Mayor}, M.,
  {Perrier}, C., {Bouchy}, F., {Lovis}, C., {Pepe}, F., {Queloz}, D., and
  {Bertaux}, J. (2007).
\newblock {The HARPS search for southern extra-solar planets. XI. Super-Earths
  (5 and 8 M+) in a 3-planet system}.
\newblock {\em Astron. Astrophys.}, 469:L43--L47.

\bibitem[Vallis, 2006]{Vallis2006}
Vallis, G.~K. (2006).
\newblock {\em Atmospheric and oceanic fluid dynamics: fundamentals and
  large-scale circulation}.
\newblock Cambridge University Press.

\bibitem[{von Paris} et~al., 2010]{vonParis2010}
{von Paris}, P., {Gebauer}, S., {Godolt}, M., {Grenfell}, J.~L., {Hedelt}, P.,
  {Kitzmann}, D., {Patzer}, A.~B.~C., {Rauer}, H., and {Stracke}, B. (2010).
\newblock {The extrasolar planet Gliese 581d: a potentially habitable planet?}
\newblock {\em Astronomy and Astrophysics}, 522:A23+.

\bibitem[Wang and Read, 2012]{Wang2012}
Wang, Y. and Read, P. (2012).
\newblock Diversity of planetary atmospheric circulations and climates in a
  simplified general circulation model.
\newblock {\em Proceedings of the International Astronomical Union},
  8(S293):297--302.

\bibitem[{Wordsworth} et~al., 2010a]{Wordsworth2010}
{Wordsworth}, R., {Forget}, F., and {Eymet}, V. (2010a).
\newblock {Infrared collision-induced and far-line absorption in dense CO2
  atmospheres}.
\newblock {\em Icarus}, 210:992--997.

\bibitem[Wordsworth et~al., 2013]{Wordsworth2013}
Wordsworth, R., Forget, F., Millour, E., Head, J., Madeleine, J.-B., and
  Charnay, B. (2013).
\newblock Global modelling of the early martian climate under a denser co2
  atmosphere: Water cycle and ice evolution.
\newblock {\em Icarus}, 222(1):1--19.

\bibitem[Wordsworth and Pierrehumbert, 2014]{Wordsworth2014}
Wordsworth, R. and Pierrehumbert, R. (2014).
\newblock Abiotic oxygen-dominated atmospheres on terrestrial habitable zone
  planets.
\newblock {\em The Astrophysical Journal Letters}, 785(2):L20.

\bibitem[{Wordsworth} et~al., 2010b]{Wordsworth2010b}
{Wordsworth}, R.~D., {Forget}, F., {Selsis}, F., {Madeleine}, J., {Millour},
  E., and {Eymet}, V. (2010b).
\newblock {Is Gliese 581d habitable? Some constraints from radiative-convective
  climate modeling}.
\newblock {\em Astronomy and Astrophysics}, 522:A22+.

\bibitem[{Wordsworth} et~al., 2011]{Wordsworth2011}
{Wordsworth}, R.~D., {Forget}, F., {Selsis}, F., {Millour}, E., {Charnay}, B.,
  and {Madeleine}, J.-B. (2011).
\newblock {Gliese 581d is the First Discovered Terrestrial-mass Exoplanet in
  the Habitable Zone}.
\newblock {\em The Astrophysical Journal Letters}, 733:L48.

\bibitem[Yang and Abbot, 2014]{Yang2014}
Yang, J. and Abbot, D.~S. (2014).
\newblock A low-order model of water vapor, clouds, and thermal emission for
  tidally locked terrestrial planets.
\newblock {\em The Astrophysical Journal}, 784(2):155.

\bibitem[Yang et~al., 2013]{Yang2013}
Yang, J., Cowan, N.~B., and Abbot, D.~S. (2013).
\newblock Stabilizing cloud feedback dramatically expands the habitable zone of
  tidally locked planets.
\newblock {\em The Astrophysical Journal Letters}, 771(2):L45.

\bibitem[Yung and DeMore, 1999]{Yung1999}
Yung, Y.~L. and DeMore, W.~B. (1999).
\newblock Photochemistry of planetary atmospheres.
\newblock In {\em Photochemistry of planetary atmospheres/Yuk L. Yung, William
  B. DeMore. New York: Oxford University Press, 1999. QB603. A85 Y86 1999},
  volume~1.

\end{thebibliography}
\bibliographystyle{apalike}

\begin{table}[h]
\centering
\caption{Standard parameters used in the main correlated-$k$ GCM simulations. Planet radius and surface gravity $r,g$ were derived from mass $M$ using the scaling relation of \cite{Sotin2007} for rocky planets.}
\begin{tabular}{ll}
\hline
\hline
Parameter & Values \\
\hline
Stellar luminosity $L$ [$L_{\Sun}$] & 0.024 \\
Stellar spectrum    & AD Leo \\
Orbital eccentricity  $e$ & 0.0 \\
Obliquity  $\phi$ & 0.0 \\
Atmospheric pressure $p$ [bar] & 0.01-10.0 \\
Stellar flux $F$ [1366~W/m$^2$] & 0.2-3.0 \\
Planet mass $M$ [$M_{\Earth}$] & 1.0, 10.0 \\
Planet radius $r$ [$r_{\Earth}$] & 1.0, 1.88 \\
Surface gravity $g$ [m~s$^{-2}$] & 9.8, 27.8 \\
Surface roughness height  $z_0$ [m] & $1\times10^{-2}$ \\
Surf. therm. inertia  $\mathcal I$ [tiu] & $250$ \\
Surface albedo  $A$ & 0.2 \\
Atmospheric composition $ $ & \ce{CO2}, \ce{CO} \\
\hline \hline
\end{tabular}\label{tab:params}
\end{table}

\begin{table}[h]
\centering
\caption{Coefficients for the empirical collapse equation (\ref{eq:pcollapse_empiric}).}
\begin{tabular}{lllll}
\hline
\hline
Planet mass & $c_1$ & $c_2$ & $c_3$  & $c_4$  \\
\hline
1.0~$M_E$ & $-0.84\times10^{-10}$ & $0.73\times10^{-6}$ & $-0.0022$  & 6.01 \\
10.0~$M_E$ & $-0.81\times10^{-10}$ & $0.63\times10^{-6}$ & $-0.0017$  & 6.34 \\
\hline \hline
\end{tabular}\label{tab:coeffs}
\end{table}

\end{document}